\documentclass{aa}

\usepackage{natbib}
\usepackage{amsmath, amsbsy, amssymb}
\bibpunct{(}{)}{;}{a}{}{,}
\usepackage{graphics}
\usepackage{graphicx}
\usepackage{epstopdf}
\usepackage{longtable}
\usepackage{url}
\usepackage{bm}
\usepackage[varg]{txfonts}
\usepackage{pdflscape}
\usepackage{supertabular}
\usepackage{hyperref} 
\hypersetup{colorlinks=true, citecolor=blue, linkcolor=blue} 
\usepackage[table, svgnames, dvipsnames]{xcolor}
\usepackage{longtable}
\usepackage{lscape}
\usepackage{booktabs}
\usepackage{multicol}
\usepackage{placeins}
\usepackage{wrapfig}
\usepackage{comment}
\usepackage{enumitem}
\usepackage{amsfonts}
\usepackage{subcaption}
\usepackage{verbatim}
\usepackage{afterpage}
\usepackage{blindtext} 
\usepackage{multirow}
\usepackage{titlesec}
\usepackage{acronym}
\usepackage{wasysym}
\usepackage{soul}
\usepackage{array}
\usepackage{textcomp} 
\usepackage{sidecap} 
\usepackage{xfrac}
\usepackage{rotating}
\usepackage{siunitx}
\usepackage{adjustbox}
\usepackage[flushleft]{threeparttable}
\usepackage{caption}
\usepackage{mathrsfs}
\usepackage{mathtools}
\usepackage{breqn}

\definecolor{mypink}{RGB}{203, 175, 192}
\definecolor{operamauve}{rgb}{0.72, 0.52, 0.65}

\def\msol{\ensuremath{M_\odot}}
\def\genec{{\sc Genec}}

\begin{document}

    \title{Fluorine production in He-burning regions of massive stars during cosmic history} 

    \author{Sophie Tsiatsiou\inst{1},
            Georges Meynet\inst{1},
            Eoin Farrell\inst{1},
            Yutaka Hirai\inst{2,3}, 
            Arthur Choplin\inst{4},
            Yves Sibony\inst{1},
            S\'ebastien Martinet\inst{4},
            Rafael Guerço\inst{5,6},
            Verne Smith\inst{7},
            Katia Cunha\inst{6,8},
            St\'ephane Goriely\inst{4},
            Marcel Arnould\inst{4},
            Jos\'e G. Fern\'andez-Trincado\inst{5},
            Sylvia Ekstr\"om\inst{1}
        }

    \institute{Department of Astronomy, University of Geneva, Chemin Pegasi 51, 1290 Versoix, Switzerland\\ \email{sofia.tsiatsiou@unige.ch} 
    \and Department of Physics and Astronomy, University of Notre Dame, 225 Nieuwland Science Hall, Notre Dame, IN 46556, USA
    \and Astronomical Institute, Tohoku University, 6-3 Aoba, Aramaki, Aoba-ku, Sendai, Miyagi 980-8578, Japan
    \and Institut d'Astronomie et d'Astrophysique, Universit\'e Libre de Bruxelles, CP 226, 1050, Brussels, Belgium
    \and Instituto de Astronom\'ia, Universidad Cat\'olica del Norte, Av. Angamos 0610, Antofagasta, Chile
    \and Observat\'orio Nacional, R. Gen. Jos\'e Cristino 77, 20921-400, Rio de Janeiro, Brazil
    \and NOIRLab, Tucson, AZ 85719 USA
    \and Steward Observatory, University of Arizona, 933 North Cherry Avenue, Tucson, AZ 85721-0065, USA
        }
    
    \date{Accepted XXX, Received YYY; in original form ZZZ}

    \abstract
    {The origin of fluorine is still a debated question. Asymptotic giant branch stars synthesise this element and likely contribute significantly to its synthesis in the present-day Universe. However, it is not clear whether other sources contribute, especially in the early Universe.}
    {We discuss variations of the surface abundances of fluorine coming from our massive star models and compare them with available present-day observations. We compute the contribution of single massive stars in producing ${}^{19}$F over metallicities covering the whole cosmic history ({\it i.e.} from zero up to super-solar metallicities).}
    {We used massive star models in the mass range of 9~{\msol} $\leq M_{\rm ini} \leq$ 300~{\msol} at metallicities from Population~III $(Z=0)$ up to super-solar $(Z=0.020)$ while accounting for the required nuclear network to follow the evolution of ${}^{19}$F during the core H- and He-burning phases. Results from models with and without rotational mixing are presented.} 
    {We find that rotating models predict a slight depletion of fluorine at their surface at the end of the main sequence phase. In more advanced evolutionary phases, only models with an initial mass larger than 25~{\msol} at metallicities $Z \geq 0.014$ show phases where the abundance of fluorine is enhanced. This occurs when the star is a Wolf-Rayet star of the WC type. WC stars can show surface abundances of fluorine ten times larger than their initial abundance. However, we obtained that the winds of massive stars at metallicities larger than $Z=0.006$ do not significantly contribute to fluorine production, confirming previous findings. In contrast, very metal-poor rapidly rotating massive star models may be important sources of fluorine through the mass expelled at the time of their supernova explosion.}
    {Observations of WC stars at solar or super-solar metallicities may provide very interesting indications on the nuclear pathways that lead to fluorine production in massive stars. The possibility of observing fluorine-rich carbon-enhanced metal-poor stars is also a way to put constrains in present models at very low metallicities.}


    \authorrunning{S. Tsiatsiou et al.}

    \maketitle

\section{Introduction} \label{sec:1}

The origin of fluorine, which has only one stable isotope (${}^{19}$F), remains a topic of discussion \citep[see e.g.][]{Ryde2020}. Various types of sources have been studied in the literature: Low and intermediate mass stars during the asymptotic giant branch (AGB) phase \citep{Mowlavi1996, Goriely2000, Kara2010, Karakas2014, Cristallo2014, Pignatari2016, Ritter2018, Battinoup2019, Battino2022}; Wolf-Rayet (WR) stars \citep{GMMA2000, Stancliffe2005, Pala2005}; rapidly rotating massive stars at low metallicity \citep{Choplin2018, Limongi2018}; supernovae (SNe) through the nucleosynthesis induced by neutrinos \citet{Woosley2002, Heger2005}; novae \citet{Jose2012}; Merging events between helium and carbon-oxygen white dwarfs \citep{Longland2011}.

There is little doubt that AGB stars play an important role in the synthesis of fluorine. Indeed, fluorine overabundances have been observed at the surface of AGB stars, in planetary nebulae, and in white dwarfs (WDs) at levels in general compatible with what is predicted by stellar models \citep{Jorissen1992, Ziu1994, Werner2005, Otsuka2008, Abia2009, Abia2010, Abia2015, AbiaCORR2015, Alves2011, Lucatello2011, Otsuka2015, Werner2015, Werner2016, Otsuka2020, Saberi2022}. Abundances of fluorine have also been obtained at the surface of other types of stars, such as main sequence (MS) dwarfs \citep{Recio2012} and red giants \citep{Cunha2003, Jonsson2017, Alves2012, Laverny2013}. These abundances either reflect the initial composition of the proto-stellar cloud that formed the star, thus providing insights into the evolution of the element's abundance in the interstellar medium (ISM), or they give insights into some mixing mechanism that occurred in the star itself. Recently \citet{RSGF2022} obtained, for the first time, the abundance of fluorine in a red supergiant (RSG) star, IRS 7, and demonstrated the potential of fluorine measurements to probe mixing processes in luminous RSGs.

At very low metallicity, fluorine has been measured at the surface of carbon-enhanced metal-poor (CEMP) stars, especially those showing surface enhancements of s-process elements, the so-called CEMP-s stars \citep[see e.g.][]{Schuler2007, Lugaro2008, Lucatello2011, Mura2020}. According to \citet{Mura2020}, some of these stars may be explained by a past accretion of matter from a now disappeared AGB companion, for instance HE 1429-0551. In other cases, such as for HE 1305+0007, the estimated upper limit of fluorine shows significant discrepancies compared to AGB models. \citet{Choplin2018} discussed the origin of overabundances of ${}^{19}$F observed in some CEMP-s, invoking rapidly rotating metal-poor massive stars.

As mentioned above, while the synthesis of fluorine in AGB stars is confirmed by observational evidence, additional sources are needed to account for the evolution of this element's abundance \citep{Abia2019}. For instance, \citet{Olive2019} found that the neutrino-process ($\nu$-process), though very uncertain, may dominate the production of ${}^{19}$F at low metallicities, whereas the production by AGB stars likely dominates at near-solar metallicities. However, when measuring the fluorine abundance in the ISM along two lines of sight using the Far Ultraviolet Spectroscopic Explorer (FUSE), \citet{Federman2005} found no clear evidence for fluorine production via the $\nu$-process in regions shaped by past SNe \citep[see also][for fluorine abundances in ISM]{Snow2007}. In contrast \citet{Li2013}, who obtained an upper limit for the fluorine abundance in seven metal-poor field giants and considered chemical evolution models with contributions from AGB stars, SNe ($\nu$-process), and WR stars concluded that models do not accurately predict the observed distribution of [F/O]. Instead, observations align better with models considering an SN Type II $\nu$-process. \citet{Renda2005} favoured production by both WR and AGB stars. Meanwhile, \citet{Spitoni2018} concluded that although the production of ${}^{19}$F is predominantly by AGB stars, a contribution from WR stellar winds is necessary to match the observed [F/O] to [O/H] ratio in the solar neighbourhood. \citet{Grisoni2020} favoured rotating massive stars at low metallicity and low-mass stars, including AGB and/or novae, at later times. Similar conclusions were obtained by \citet{Womack2023}. Keeping in mind that the production of fluorine in metal-poor rapidly rotating massive stars can occur in Population (Pop) III stars and thus has properties of a primary channel of production ({\it i.e.} is produced by stars even completely deprived of any enrichment in heavy elements by a previous generation of stars), this would support the view by \citet{Guer2019}, who found that at low metallicities ([Fe/H] $< -0.4$ to $-0.5$), the abundance of fluorine varies as a primary element with respect to iron abundance, showing a plateau around [F/Fe] $\sim-0.3$ to $-0.4$. However, at higher metallicities, [F/Fe] increases with [Fe/H], indicating a near-secondary behaviour. In line with these findings, \citet{Roberti2024a} show (see their Figs.~17 and 18) that ${}^{19}$F is primarily produced during shell He-burning — particularly in fast rotators at low metallicities — when the helium convective shell develops in the region containing the so-called CNO pocket. However, as also demonstrated in \citet{Roberti2024b}, an extension of the \citet{Limongi2018} grids to super-solar metallicities, this entanglement effect is strongly suppressed at high metallicity due to enhanced mass loss.

The above discussion shows that, at the moment, contradictory conclusions have been reached on the history of fluorine production in the Universe, thus justifying the exploration of different stellar models. This is the main aim of this work.

In this paper, we discuss three aspects related to the nucleosynthesis of fluorine in massive stars. As mentioned above, AGB stars may not be the only source of fluorine. In this work, first, we provide predictions for fluorine abundances on the surface of massive stars of different masses, metallicities, and rotation rates and at various evolutionary stages. Our goal is to find stars that could help probe the production process of fluorine. Then, we re-examine whether stellar winds play a significant role in ejecting fluorine-enriched material based on our latest grids of massive star models. Finally, using those models, we explore whether the production of fluorine is boosted in very metal-poor rotating massive stars, known as well as spinstars. These two questions have already been explored using {\genec} stellar models \citep[][]{GMMA2000,Palacios2005,Choplin2018} but for different mass domains, metallicities, overshooting, initial rotations, diffusion coefficients for the mixing of the elements by rotation and the mass-loss rates, and the dependence of the mass-loss rates on the metallicity (as seen in Table~{\ref{table:models}). We note that for each reference and for the models with rotation, the first line gives the reference of the vertical shear diffusion coefficient, and the second line provides the reference for the horizontal shear diffusion. Similarly, the mass-loss rate recipe used during most of the pre-WR phase is indicated on the first line, the second gives the reference for the WR stellar winds, and the third line provides the metallicity dependence of the stellar winds. In this work, a different metallicity dependence has been used for the pre-WR and the WR phase. More details about the physical ingredients of these models are given in the references.

The organisation of the paper is as follows: In Sect.~\ref{sec:2}, we describe the ingredients of the stellar models. The predictions of the present models for fluorine abundances on the surface of massive stars, including RSG and WR stars are presented in Sect.~\ref{sec:3}. In Sect.~\ref{sec:4}, we investigate whether stellar winds can significantly contribute to fluorine at high metallicities. Sect.~\ref{sec:5} is devoted to the production of fluorine by spinstars. The stellar yields obtained from models at different metallicities are presented in Sect.~\ref{sec:6} (we use a very simple chemical evolution model). Finally, the conclusions are provided in Sect.~\ref{sec:7}.

    \begin{table*}
    \small
    \caption{Comparisons between the physical ingredients of {\genec} models providing fluorine stellar yields.}
    \begin{center}
    \begin{tabular}{cccccccc}
    \hline									
    \hline \noalign{\smallskip}										
    Reference        & Mass domain	& $Z$	   & $d_{\rm over}/H_P$ &       $\upsilon_{\rm ini}$    & Diffusion coefficients:   & Mass loss:  	\\
                     &    &          & 	              &                                & $D_{\rm shear}$ & pre-WR    \\
                     &       	   &          & 	              &                                & $D_{\rm h}$     &  WR      	\\
                     &              &          &                   &                                &                 &  Z-dependence           \\
               
    \hline \noalign{\smallskip}								
    \citet{GMMA2000}    & $25-120$~{\msol} & $0.008-0.040$ & 0.20             &      $0\,{\rm km}\,{\rm s}^{-1}$                        & --                & \citet{deJager1988}  \\
                        &         &             &                  &                                & --                & \citet{Conti1988}                \\
                        &         &             &                  &                                &                   & $(Z/Z_\odot)^{0.5}$ \\
    \citet{Pala2005}    & $25-120$~{\msol} & $0.004-0.040$ & 0.10             & $0-300\,{\rm km}\,{\rm s}^{-1}$               & \citet{Talon1997} & \citet{Vink2001}      \\
                        &         &             &                  &                                & \citet{Zahn1992}  & \citet{Nugis2000}      \\
                        &         &             &                  &                                &                   & $(Z/Z_\odot)^{0.5}$ \\                      
    \citet{Choplin2018} & $10-150$~{\msol} & $10^{-5}$   &          0.10        & $0.4\,\upsilon_{\rm crit}$     & \citet{Talon1997} & \citet{Vink2001}         \\
                        &         &             &                  &                           $(214-490\,{\rm km}\,{\rm s}^{-1})$    &  \citet{Zahn1992} & --                       \\
    This work           & $9-300$~{\msol}  & $0-0.020$   & 0.10             & $0.4\,\upsilon_{\rm crit}$     & \citet{Maeder1997}& \citet{Vink2001}          \\
                        &         &             &                  &                           $(225-708\,{\rm km}\,{\rm s}^{-1})$     & \citet{Zahn1992}  & \citet{Nugis2000}       \\                     
                        &         &             &                  &                                &                   & $(Z/Z_\odot)^{0.66-0.85}$ \\       
    \hline
    \end{tabular}
    \end{center}
    \label{table:models}
    \end{table*}

\section{Stellar models} \label{sec:2}

\subsection{Ingredients of the stellar models} \label{sec:21}

The stellar models used in this study are derived from several sources. \citet{Murphy2021} and \citet{Tsiatsiou2024} provided models for Pop~III at $Z=0$. Extremely metal-poor (EMP) models at $Z=10^{-5}$ come from \citet{Sibony2024}, while IZw18 galaxy metallicity models at $Z=0.0004$ from \citet{Groh2019}. Models with Small Magellanic Cloud (SMC) metallicity at $Z=0.002$ were developed by \citet{Georgy2013a}, and those with Large Magellanic Cloud (LMC) metallicity at $Z=0.006$ by \citet{Eggenberger2021}. \citet{Ekstrom2012} developed solar metallicity models at $Z=0.014$, and \citet{Yusof2022} super-solar metallicity models at $Z=0.020$. 

All these model grids were computed using the Geneva stellar evolution code \citep[{\genec}, see a detailed description in][]{Eggenberger2008}. They share the same input physics, including the treatment of convection (Schwarzshild criterion for convection with a step overshoot equal to $0.1\,H_P$, where $H_P$ is the pressure scale height measured at the Schwarzschild boundary of the convective core), mass loss, rotation (shellular rotating models accounting for shear turbulence and meridional transport process), nuclear reaction rates (see below), and opacities\footnote{For a detailed description of the physical ingredients of the models see \citet{Ekstrom2012}.}. This ensures that variations in model outputs at different metallicities arise solely from changes in metallicity. These models thus provide a homogeneous set of stellar models. As with all models, they have their strengths and weaknesses, and some of them are discussed briefly in Sect.~\ref{sec:7}. 

Models by \citet{Martinet2023} for very massive stars (VMSs), with masses ranging between 180 and 300~{\msol}, are used for $Z=0$, $Z=10^{-5}$, $Z=0.006$, and $Z=0.014$ metallicities. These models have slightly different input physics compared to the others. Specifically, they adopt the Ledoux criterion for convection with a step overshoot, where the convective radius at the Ledoux boundary is increased by an amount $\Delta R_{\rm Led}=0.2\,H_P$. This change of the criterion for convection and for the overshooting parameter for these VMSs was triggered by two studies: first, \citet{Georgy2014} concluded that for massive stars, Ledoux criterion could better reproduce the nitrogen enrichment at the surface of blue supergiants (BSG) having evolved from a RSG stage; second, \citet{Martinet2021} studied the impact of different overshootings on the MS width and concluded that the overshoot should increase with the initial mass of the star. Although these models indeed differ by their physics, we still accounted for their predictions here because those stars have anyway very large convective core and thus approach a near homogeneous evolution for a very large part of their mass. Thus, their outputs are less sensitive to the changed of convective criterion and of the overshooting parameter.

The massive rotating Pop~III models have initial surface equatorial velocities equal to either 40\% or 70\% of the critical one\footnote{The critical velocity is defined as the rotation velocity at the equator such that the sum of the centrifugal and radiative forces balance the gravity there. We use the formulation of the critical velocity given by \citet{OG2000}.} at the zero-age main sequence (ZAMS). Hereafter, these models are referred to as moderately rotating or rapidly rotating models, respectively. The rotating models for all the other metallicities correspond to an initial velocity equal to 40\% of the critical one.

The same chains of reactions as in \citet{GMMA2000} are used. We repeat them below to make the paper more self-consistent. The reaction chain important for following the variations of the fluorine abundances during the core H-burning phase is
$$^{14}{\rm N}(p,\gamma)^{15}{\rm O}(\beta^+)^{15}{\rm N}(p,\gamma)^{16}{\rm O}(p,\gamma)^{17}{\rm F},$$
$$^{17}{\rm F}(\beta^+)^{17}{\rm O}(p,\gamma)^{18}{\rm F}(\beta^+) ^{18}{\rm O}(p,\gamma) ^{19}{\rm F}(p,\alpha)^{16}{\rm O}.$$
\noindent In general, fluorine is destroyed in the core of massive stars during the core H-burning phase. The reaction chain important for following the variations of the fluorine abundance during the core He-burning phase is
\begin{multline*}
{}^{14}{\rm N}(n,p){}^{14}{\rm C}(\alpha,\gamma){}^{18}{\rm O}(p,\alpha){}^{15}{\rm N} \\
\end{multline*}
\vspace{-1.2cm}
\begin{eqnarray*}
                            &             & \hspace{1.7cm}\raisebox{1ex}{$\searrow$} \cr 
                            &             &
(\beta^+)^{18}{\rm O}(p,\alpha)^{15}{\rm N}(\alpha,\gamma)^{19}{\rm F}                \cr
                            & \nearrow    &  \hskip 3.8cm  \searrow            
                                  \cr 
^{14}{\rm N}(\alpha,\gamma)^{18}{\rm F} & \rightarrow &
(n,p)^{18}{\rm O}(p,\alpha)^{15}{\rm N}(\alpha,\gamma)^{19}{\rm F}(\alpha,p)^{22}{\rm Ne}  \cr
                            & \searrow    &   \hskip 3.8cm  \nearrow           
                                  \cr
                            &             & \hskip 0.5cm
(n,\alpha)^{15}{\rm N}(\alpha,\gamma)^{19}{\rm F}.                               \cr
\end{eqnarray*}
\noindent We see that the fluorine synthesis in He-burning regions starts from $^{14}$N. This $^{14}$N originates from the operation of the CNO cycle in H-burning regions. The synthesis of $^{19}$F occurs as one possible channel of alpha or neutron captures by the $^{14}$N left by the previous CNO cycle. It is well known that, at the beginning of the core He-burning phase, most of $^{14}$N is transformed into $^{22}$Ne, but, through the channel presented just above, a fraction of it can be transformed into $^{19}$F. We note that fluorine production requires the availability of neutrons and/or protons. They are mainly produced by the reactions $^{13}{\rm C}(\alpha,n)^{16}$O and $^{14}{\rm N}(n,p)^{14}$C.

The nuclear reaction rates in all the different grids of models at different metallicities are the same as in \citet{Ekstrom2012}\footnote{The rates for the reactions involving fluorine are from the NACRE compilation \citep{Angulo1999NACRE}. The NACRE~II update \citep{Xu2013} was published after the solar models of the present series of models were computed \citep{Ekstrom2012}.}. These rates have not been updated in order to keep the same physics in all the grids. This will also allow for a comparison with the results of \citet{Pala2005}, who used the same code with the same nuclear reaction rates but different prescriptions for the mass loss and the rotational mixing. 

\subsection{Stellar yields computation} \label{sec:22}

Stellar wind yields ($p_{19}^{\rm wind}$) are the masses of newly synthesised ${}^{19}$F ejected by stellar winds by a star of a given initial mass, metallicity, rotation. It is given by
\begin{dmath}
    {p_{19}^{\rm wind}(M_{\rm ini},Z,\upsilon_{\rm ini})=} \\
    \int_{\:0}^{\:t(M_{\rm ini},Z,\upsilon_{\rm ini})} \;\;\; \dot{M}(t', M_{\rm ini},Z,\upsilon_{\rm ini}) \;\;\; {\left(^{19}{\rm F}_{\rm s}(t', M_{\rm ini},Z,\upsilon_{\rm ini})-{}^{19}{\rm F}_{\rm ini} \right) {\rm d}t'}
\end{dmath}
where $\dot{M}$ represents the mass lost per unit time in solar masses by stellar winds when a star of initial mass $M_{\rm ini}$, initial metallicity $Z$, and initial equatorial rate $\upsilon_{\rm ini}$ ($\upsilon_{\rm ini}/\upsilon_{\rm crit}$ on the ZAMS) reaches the age $t'$. $^{19}{\rm F}_{\rm s}$ is the surface mass fraction of fluorine, ${}^{19}{\rm F}_{\rm ini}$ is the initial fluorine abundance. The values are negative when the mass of ejected fluorine by the winds is lower than the mass that was initially present in that same portion of the star.

The SNe yields ($p_{19}^{\rm SN}$) are the masses of newly produced fluorine ejected at the time of the SN explosion. They are computed using the following formula:
\begin{multline} 
    p_{19}^{\rm SN}(M_{\rm ini},Z)= \\ 
     = \int_{M_{\rm cut}}^{M_{\rm preSN}} \left[{}^{19}{\rm F}(t_{\rm preSN}, M_{\rm r}, M_{\rm ini}, Z)-{}^{19}{\rm F}_{\rm ini} \right]{\rm d}M_{\rm r},
\end{multline}
where $M_{\rm cut}$ is the remnant mass (i.e. the mass of the central part of the SN progenitor that remains locked within the remnant), $M_{\rm preSN}$ is the actual mass of the star at the SN stage, and ${}^{19}{\rm F}(t_{\rm preSN}, M_{\rm r}, M_{\rm ini}, Z)$ represents the mass fraction of fluorine at the lagrangian mass coordinate $M_{\rm r}$ in the model with an initial mass $M_{\rm ini}$ and initial metallicity $Z$ at the age $t_{\rm preSN}$ of the SN. The method proposed by \citet{Patton2020} is followed to compute the remnant mass. In general, the remnant mass is well below the zone of the star that has been enriched in fluorine by He-burning. The yields do not include contributions from explosive nucleosynthesis.

The total stellar yields ($p_{19}^{\rm tot}$) of one star is the sum of the wind and SN yields, that is, $p_{19}^{\rm tot}=p_{19}^{\rm wind}+p_{19}^{\rm SN}$.

\section{Predictions for fluorine surface abundances} \label{sec:3}

Abundances of fluorine at the surface of stars can provide insights on the nuclear transformation of this element inside the star. For instance, when material processed by H-burning reactions is exposed at the stellar surface, one expects to observe a depletion of fluorine, since fluorine is mainly destroyed during this phase by proton capture. Instead, when material transformed by He-burning appears at the surface, one expects either an overabundance of fluorine (with respect to the initial abundance) or a depletion, as further detailed below. 

Some stars, such as AGB stars, exhibit a mixture between H- and He-burning products. In general, overabundances have been observed at the surface of AGB stars \citep[see e.g.][]{Jorissen1992, Saberi2022}, supporting the view that these stars are important sources of fluorine. In this section, we shall not discuss the case of AGB stars \citep[see e.g. models by][]{Kara2010, Karakas2014, Cristallo2014, Pignatari2016, Ritter2018, Battinoup2019, Battino2022}, but we focus on massive stars. More precisely, we address the following questions:
\begin{enumerate}
    \item Can rotational mixing, together with mass loss, impact the surface fluorine abundance during the core H-burning phase of massive stars?
    \item What are the abundances expected during the RSG phase?
    \item What are the fluorine abundances expected in stripped stars (helium-rich stars)?
\end{enumerate}

\subsection{Fluorine surface abundance for stellar models with masses between 12 and 25~{\msol}} \label{sec:31}

    \begin{figure}
    \includegraphics[width=0.5\textwidth]{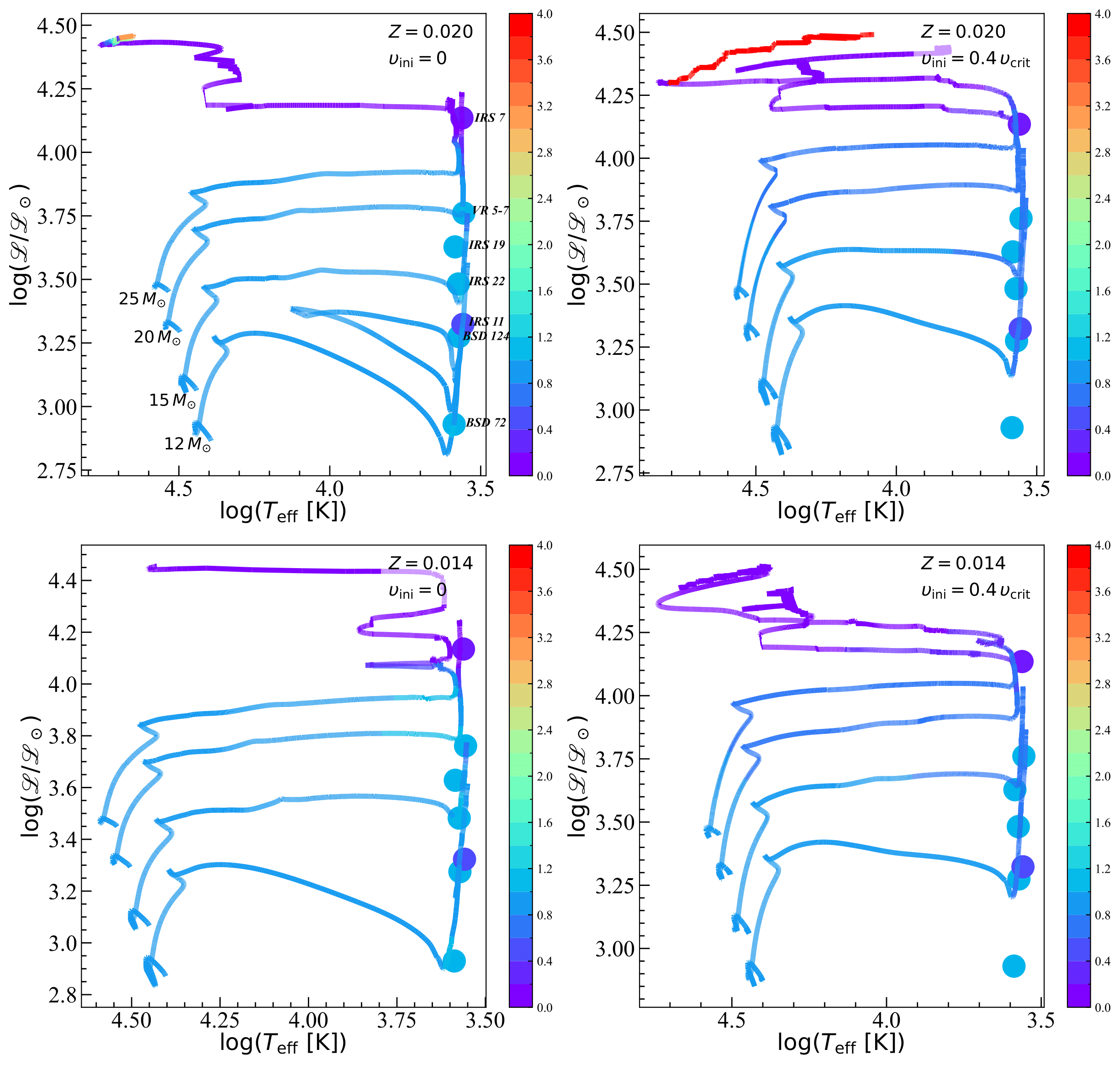}   
    \caption{Evolutionary tracks in the spectroscopic HR diagram (where the logarithm of the luminosity is replaced by the logarithm of ${\cal L}=T_{\rm eff}^4/g$, with $g$ being the surface gravity). The colour scale indicates the surface value of N(${}^{19}$F$_{\rm s}$)/N($^{1}$H$_{\rm s}$), normalised to the ratio's initial value at ZAMS, where N represents the number density of the considered isotope. The rotation and the metallicity are indicated in each panel. The points represent the fluorine abundance determined from observations by \citet{Guer2022} and \citet{RSGF2022}.}
    \label{fig:HRD} 
    \end{figure} 

The colours along the tracks shown in Fig.~\ref{fig:HRD} show how the number ratio of ${}^{19}$F to $^{1}$H evolves at the surface during the evolution of massive star models. First, let us discuss the non-rotating models at $Z=0.020$ (upper left panel). We observe that throughout the entire MS phase and during the first crossing of the Hertzsprung-Russel (HR) gap, the surface ratio of fluorine to hydrogen remains equal to its initial value. Then, during the ascent of the RSG branch, the ratio decreases. It becomes lower than 0.3 at the surface of the 20 and 25~{\msol} models when ${\cal L} > 4.0$ (${\cal L}=T^4_{\rm eff}/g$, where $T_{\rm eff}$ is the effective temperature and $g$ the effective gravity). Only the 25~{\msol} model experiences sufficiently strong mass loss to uncover layers that have been enriched in fluorine (as indicated by the number ratio higher than 1.0 when $\log(T_{\rm eff}) \sim 4.7$ at the end of the evolution). In this more advanced stages the surface begins to be enriched in He-burning products.

The non-rotating models at $Z=0.014$ (lower left panel) show qualitatively similar behaviour. The main difference is that these models never show an increase in the surface fluorine abundance due to weaker stellar winds at a given position in the HR diagram.

Rotating models (right panels) of 20 and 25~{\msol} show decrease of the surface fluorine already during the MS phase as a result of the internal mixing triggered by rotation. This mixing shifts at slightly higher values of ${\cal L}$, the region of the spectroscopic HR diagram where depletion below 0.3 is predicted. It occurs at ${\cal L} \sim 4.0$ in the non-rotating models at $Z=0.020$, while in the same metallicity rotating models it occurs at ${\cal L} \sim 4.2$. The 25~{\msol} track in the rotating models at $Z=0.020$ shows a large domain in the spectroscopic diagram where fluorine is enriched by more than a factor of 3. The rotating models at $Z=0.014$ show a similar behaviour, though they never show fluorine surface enrichment because of weaker mass loss. 

We conclude that, in the mass range below 20~{\msol}, the surface fluorine abundances hardly change during the MS phase and the first crossing of the HR gap, for the two initial rotations considered here. Only for ${\cal L} > 4.0$, a decrease in fluorine surface abundance by more than a factor of 2.5 can be observed. Fluorine enrichment is predicted at the end of the evolution of 25~{\msol} models at $Z=0.020$ with and without rotation.

In Fig.~\ref{fig:HRD} we include observational data by \citet{Guer2022} and \citet{RSGF2022}. Most of the stars have fluorine surface abundances that are compatible with no change with respect to the initial value (see the points with a colour correspond to value equal 1). This can be seen also in Fig.~5 of \citet{RSGF2022}. Two stars present fluorine depletion. One IRS 11 is only at two sigmas below the level of initial fluorine abundance. On the other hand IRS 7 is clearly depleted in fluorine being at about $6\sigma$ below the plateau representative of the initial value. This level of depletion is compatible with the value predicted by the non-rotating 25~{\msol} model at $Z=0.020$ and $Z=0.014$. The current moderately rotating models fail by a narrow margin to reproduce the depletion observed in IRS 7 at its observed position in the spectroscopic diagram. However when one accounts for the error bar on the observed position, we conclude that the rotating models predictions are still compatible with the observed values of IRS 7\footnote{On Fig.~\ref{fig:HRD}, the error bar along the x-axis is small, around $\pm$0.018~dex, the error bar along the y-axis is much larger of the order of $\pm$0.3~dex, using errors indicated in Table~2 of \citet{Guer2022}.}.

\subsection{Fluorine surface abundance for stellar models with masses larger than 25~{\msol}} \label{sec:32}

    \begin{figure*}
    \centering
    \includegraphics[width=0.9\textwidth]{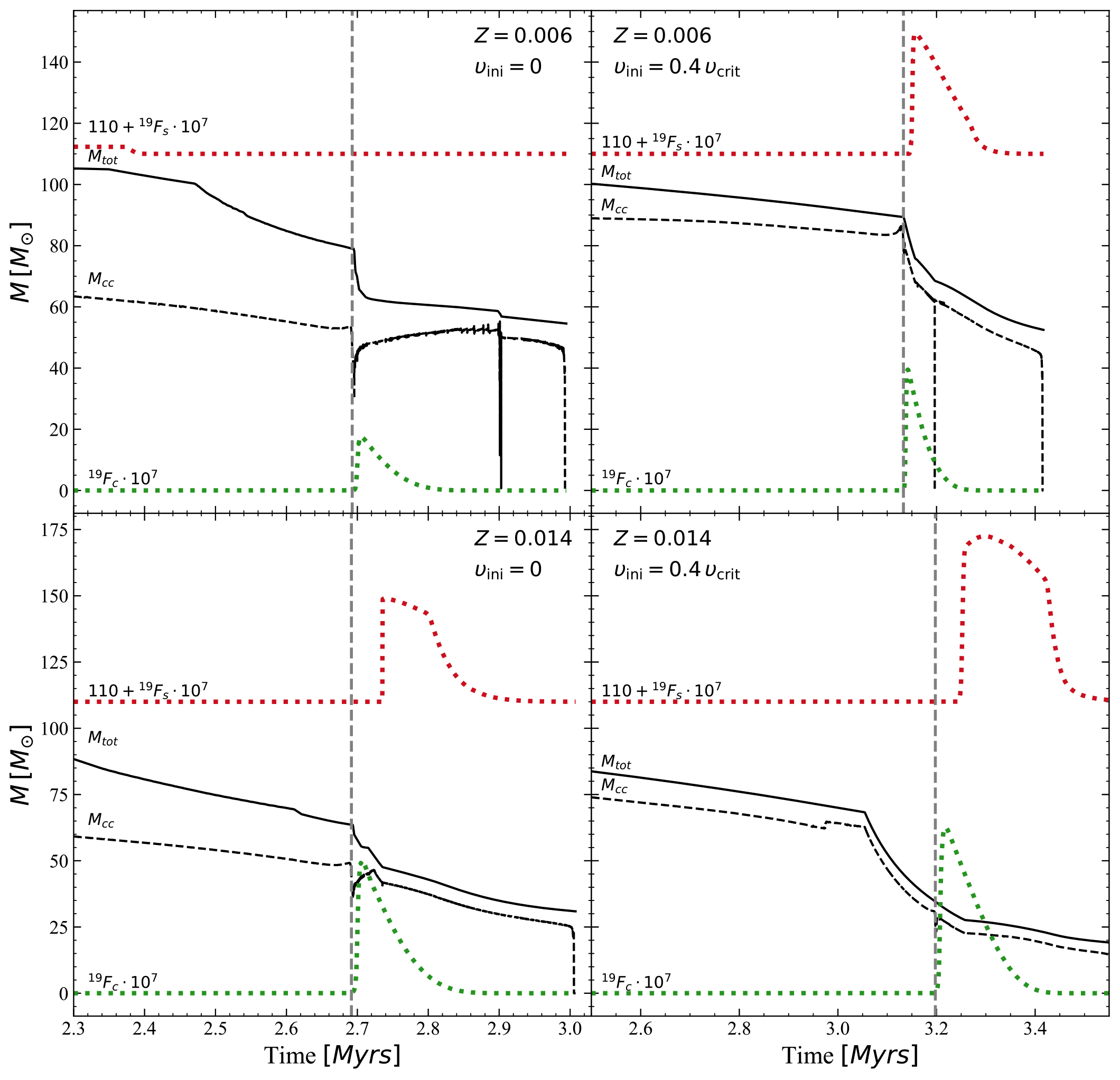}
    \caption{Evolution of $^{19}{\rm F}$ at the surface (${\rm F}_{\rm s}$) and at the centre (${\rm F}_{\rm c}$) of a 120~{\msol} star for different metallicities and rotations, as indicated in each panel (in arbitrary units). The evolution of the total mass ($M_{\rm tot}$) and the mass of the convective core ($M_{\rm cc}$) are also shown. The gray dashed lines indicate the end of the core H-burning phases. The upper panels show models at LMC metallicity, while the lower panels show models at solar metallicity.}
    \label{fig:WRF19}
    \end{figure*}   
    
    \begin{figure*}
    \centering
    \includegraphics[width=0.9\textwidth]{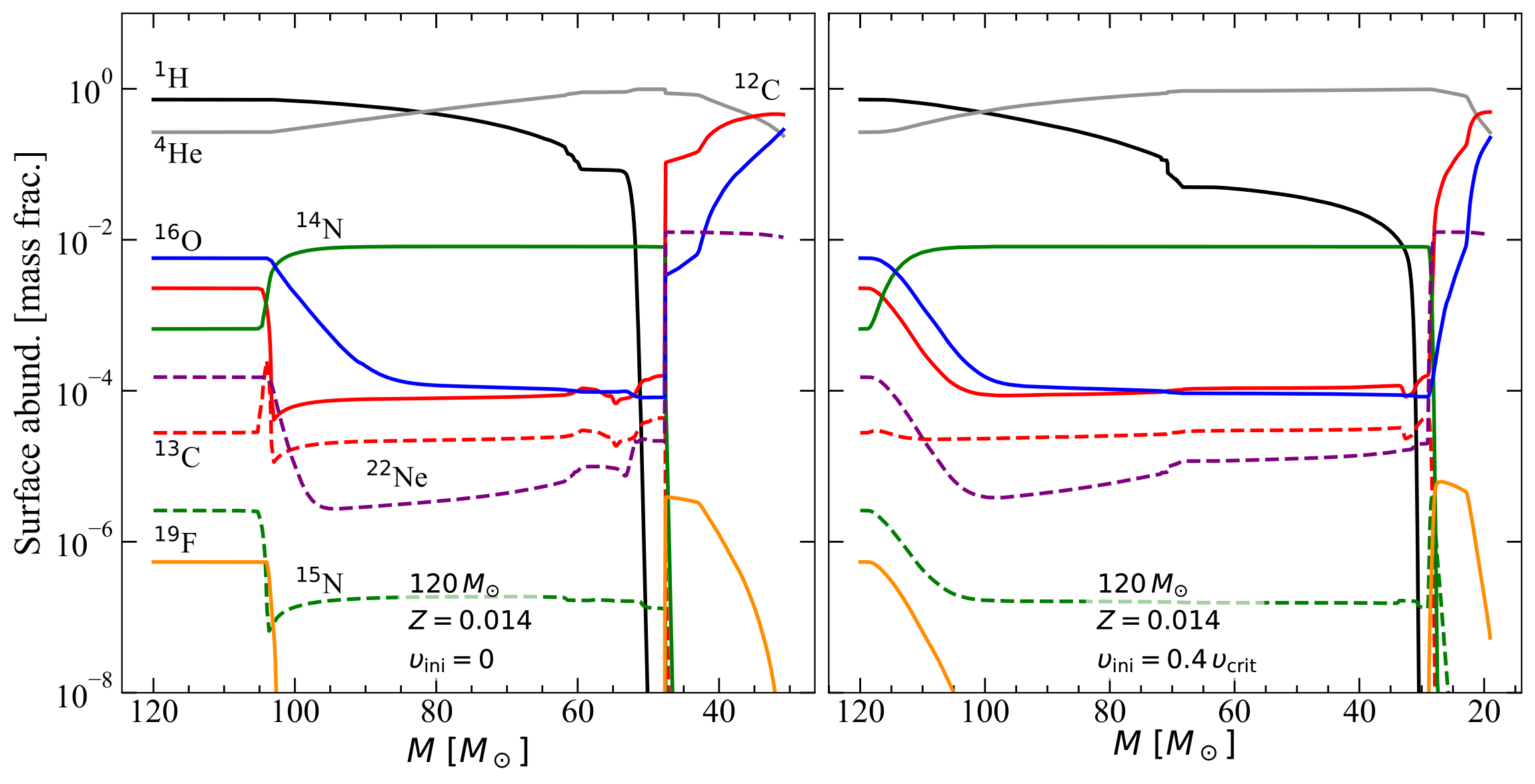}  
    \caption{Evolution of the surface abundances of a few isotopes as a function of the actual mass of a 120~{\msol} model at solar metallicity throughout its evolution (meaning as time increases, due to mass loss, the stellar mass decreases). The left panel shows the results of the non-rotating model, while the right panel shows the results of the moderately rotating model.}
    \label{fig:120Mabund}
    \end{figure*}     

    \begin{figure*}
    \centering
    \includegraphics[width=1\textwidth]{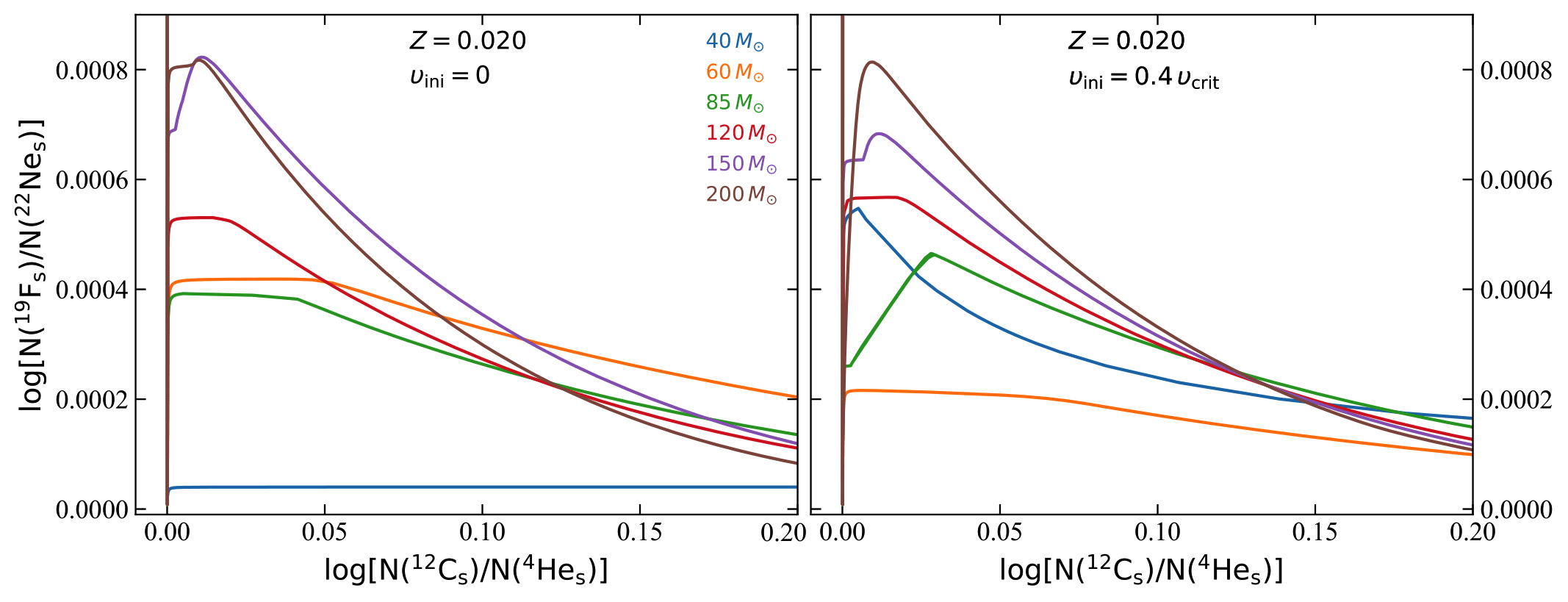}  
    \caption{Evolution of the surface abundances of the number ratio $^{19}$F to $^{22}$Ne as a function of the number ratio $^{12}{\rm C}$ to $^{4}{\rm He}$ for different initial mass stars at super-solar metallicity. The left panel shows the results from non-rotating models, while the right panel shows the results of moderately rotating models.}
    \label{fig:F19Z20S04}
    \end{figure*}        

    \begin{figure*}
    \centering
    \includegraphics[width=1\textwidth]{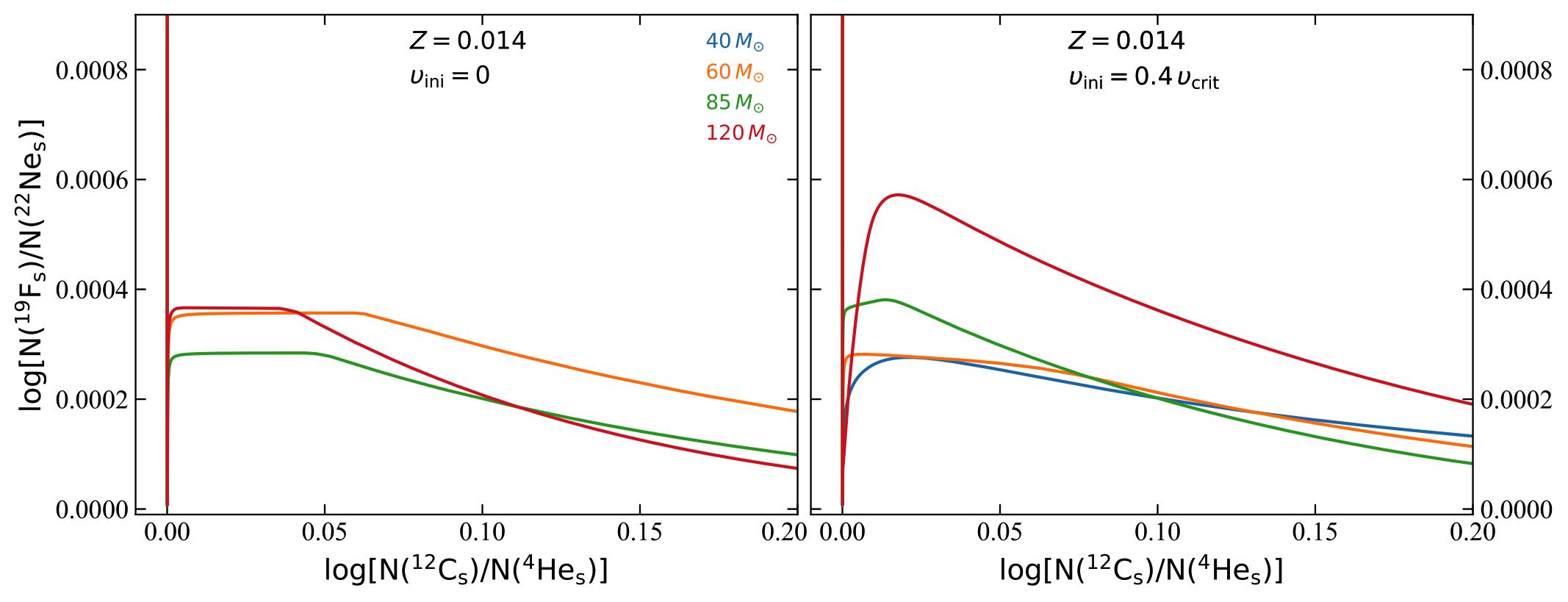}  
    \caption{Same as Fig.~\ref{fig:F19Z20S04} but for models at solar metallicity.}
    \label{fig:F19Z14S04}
    \end{figure*}            

Figure~\ref{fig:WRF19} shows the evolution of the total mass ($M_{\rm tot}$) and of the convective cores ($M_{\rm cc}$) during the H- and He-burning phases for 120~{\msol} models at different metallicities and rotations, with the gray vertical dashed lines indicating the end of the core H-burning phase. It also shows the evolution of the central ($^{19}{\rm F}_{\rm c}$) and surface ($^{19}{\rm F}_{\rm s}$) fluorine abundances (for purpose of clarity, the abundances of fluorine have been multiplied by a factor of $10^7$ and the curves for the central and surface abundances have been shifted as indicated on the figure). 

At the $Z=0.006$ metallicity for the non-rotating model (left upper panel), we can see that due to mass loss, which uncovers layers in which fluorine has been partially destroyed, the surface abundance decreases at an age of 2.36~Myrs (we note that the x-axis begins at 2.30~Myrs, thus showing only the end of the core H-burning phase). At an age of 2.69~Myrs, when the star enters the core He-burning phase, a strong increase (by a factor of 17) in the fluorine abundance can be observed at the centre, but it is quickly destroyed. This is the consequence of the nuclear sequence proposed by \citet{Goriely1989} (see Sect.~\ref{sec:2}), which is activated just at the beginning of the core He-burning phase. This reaction chain relies on free neutrons, which are released in the He-burning core. For example, in a 20~{\msol} model at $Z=10^{-5}$ metallicity with moderate initial rotation, the maximum neutron densities reached during the core and shell He-burning phases at $T \sim 10^7$~K and $\rho \sim 10^3~{\rm g}\,{\rm cm}^{-3}$, and at $T \sim 10^8$~K and $\rho \sim 10^3~{\rm g}\,{\rm cm}^{-3}$, are typically around $n_n \sim 10^6~{\rm n}\,{\rm cm}^{-3}$ and $n_n \sim 10^5~{\rm n}\,{\rm cm}^{-3}$, respectively. These conditions facilitate s-process nucleosynthesis. This process may also influence fluorine synthesis, as evidenced by the correlation between neutron densities and fluorine production. Future studies comparing the predicted neutron densities in the core and shell He-burning phases with observed s-process element abundances could provide additional constraints on these nucleosynthetic pathways. In this specific model, this increase in the centre never impacts the surface abundances, since the mass losses do not uncover layers that have been enriched in fluorine. We see that at the moment when the peak fluorine abundance is reached, the mass of the He-burning core is about 45~{\msol}, well embedded inside the star, which at that same age has a mass equal of 65~{\msol}. 

Thus, as previously discussed by \citet{Palacios2005}, stellar winds may eject substantial amounts of fluorine only if the He-core is uncovered at an early phase of the core He-burning phase and if the mass-loss rates are important at that moment. Indeed, efficient ejection of ${}^{19}$F occurs only if layers displaying the characteristic abundances of the very beginning of the He-burning phase are exposed at the surface. This exposition may be favoured by convection, rotational mixing and by the mass losses. 

For the moderately rotating model of Fig.~\ref{fig:WRF19} (right upper panel), the effects of rotational mixing on the predictions are shown. In this case, fluorine-rich material appears at the surface. We observe that mass losses become very important just before the surface becomes fluorine-rich, as the total mass of the star decreases rapidly. This corresponds to a phase when the star loses a lot of mass and enters a Luminous Blue Variable (LBV) phase. Very little time shift is observed between the peak of fluorine reached at the centre and the one occurring at the surface. This is primarily an effect of mass loss rather than of rotational mixing. Fluorine is destroyed by $\alpha$-captures at the centre. As layers affected by the later stages of the core He-burning phase, where fluorine is depleted, become exposed, the surface abundance of fluorine decreases. Thus, both at the surface and at the centre, the abundance rapidly decreases after the peak. 

At $Z=0.014$ metallicity (lower panels), the stronger mass-loss rates during the MS phase uncover the He-burning core at earlier phases than at lower metallicities. This favours strong fluorine surface enrichment. Rotation further amplifies this trend. These results suggest the existence of fluorine-rich He-stripped stars.

This can also be seen in Fig.~\ref{fig:120Mabund} for the 120~{\msol} non-rotating model with $Z=0.014$ metallicity (left panel). We see that the fluorine (orange curve) decreases from its initial abundance value ($5.4\times10^{-6}$ in mass fraction) to a value below $10^{-8}$ in mass fraction when the mass of the model is around 102~{\msol}. This decrease in surface fluorine abundance occurs when the stellar surface becomes nitrogen-rich, {\it i.e.} when CNO-processed material appears at the surface. 

When the surface is enriched with He-burning products, for instance when the actual mass of the star is around 48~{\msol}, there is a phase where the surface fluorine abundance increases by a factor of 7 from its initial value. This appears when the surface is strongly enriched in carbon and oxygen, which occurs when the star enters the WC phase\footnote{Carbon sequence WR}. At this point, fluorine is strongly enhanced at the surface. The star is expected to lose about 5~{\msol}, indicating that the mass of fluorine ejected by winds during that specific phase will be on the order of $2.7\times10^{-5}$~{\msol}. Although at that phase, the wind is loaded with substantial amounts of fluorine, at other phases, the wind is depleted in fluorine. The wind stellar yields discussed in the next section accounts for the effects of the winds over the whole stellar lifetime. In this specific case, more material is ejected with depleted fluorine than enriched in fluorine, and thus the net effect will be that this star will actually eject depleted fluorine material (see Table~\ref{table:yieldswindsF19}). The right panel of Fig.~\ref{fig:120Mabund} illustrates the same model with rotational mixing accounted for. We see that the surface abundance of fluorine decreases more gradually than in the non-rotating model. This reflects the fact that mixing induced by rotation smooths the internal chemical composition gradients, which are then revealed when mass loss peels off the outer layers of the star. Since rotation increases the MS lifetime, the star enters the core He-burning phase with a smaller actual mass. Consequently, the surface becomes enriched in He-burning products at smaller mass. On the whole however, rotation has not a very important effect on the wind ejection of fluorine in that model.

Figures~\ref{fig:F19Z20S04} and \ref{fig:F19Z14S04} show the evolution of the fluorine abundance at the surface of WC stars for $Z=0.020$ and $Z=0.014$ models, respectively. As horizontal axis, we chose the number ratio N(${}^{12}$C$_{\rm s}$)/N($^{4}$He$_{\rm s}$) since when the star enters the WC phase, this ratio increases from values approaching zero in the domain covered by the horizontal axis of Fig.~\ref{fig:F19Z20S04} and \ref{fig:F19Z14S04} up to 0.8 at maximum. The vertical axis is the surface number ratio N(${}^{19}$F$_{\rm s}$)/N($^{22}$Ne$_{\rm s}$). During the WC phase, the abundance of ${}^{22}$Ne at the surface remains nearly constant. The evolution of the ratio therefore follows the variation of ${}^{19}$F at the surface\footnote{The level of abundance of $^{22}$Ne at the surface during the WC phase corresponds nearly perfectly to the initial abundance of CNO elements in the models. This is because ${}^{22}$Ne results from the transformation, at the beginning of the core He-burning phase, of the $^{14}$N left by the CNO process in the core at the end of the core H-burning phase, this nitrogen itself resulting from the transformation of the carbon and oxygen initially present in the star. \citet{Limongi2018} have defined a parameter $\chi$ that scales with the sum of the number fraction of isotopes between $^{14}$N and $^{26}$Mg that keeps a constant value when there is no interactions between H- and He-burning zones and increases when interactions occur \citep[see Eq.~3 in][]{Roberti2024a}.}.

Provided the He-core is exposed at the beginning of the core He-burning phase, an increase in the N(${}^{19}$F$_{\rm s}$)/N($^{22}$Ne$_{\rm s}$) ratio should be observed, as demonstrated in Fig.~\ref{fig:F19Z20S04} and \ref{fig:F19Z14S04}. The vertical line at $x=0$ shows the evolution of this ratio during phases preceding the onset of the WC phase. Initially, in Fig.~\ref{fig:F19Z20S04} and \ref{fig:F19Z14S04}, the ratio attains a higher value (specifically, equal to 0.004 for the metallicities shown). This ratio then declines vertically as the N($^{12}$C$_{\rm s}$)/N($^{4}$He$_{\rm s}$) ratio consistently remains well below 1 prior to the WC phase. It is only with the emergence of He-burning products at the surface that significant increases in the abundances of carbon, fluorine, and neon are observed. The curve subsequently reaches a maximum value before showing a decrease, progressing towards higher values of N(${}^{12}$C$_{\rm s}$)/N($^{4}$He$_{\rm s}$).

Looking at Figs.~\ref{fig:F19Z20S04} and \ref{fig:F19Z14S04}, we see that the maximum values are reached for N($^{12}$C$_{\rm s}$)/N($^{4}$He$_{\rm s}$) values of a few percent. Also, the maximum value in general increases with the mass. No big differences are obtained comparing non-rotating and moderately rotating models at $Z=0.020$. 
At $Z=0.014$. since the mass-loss rates are weaker, we see more effects of rotation.

\section{${}^{19}$F wind ejection} \label{sec:4}

    \begin{figure}
    \includegraphics[width=0.5\textwidth]{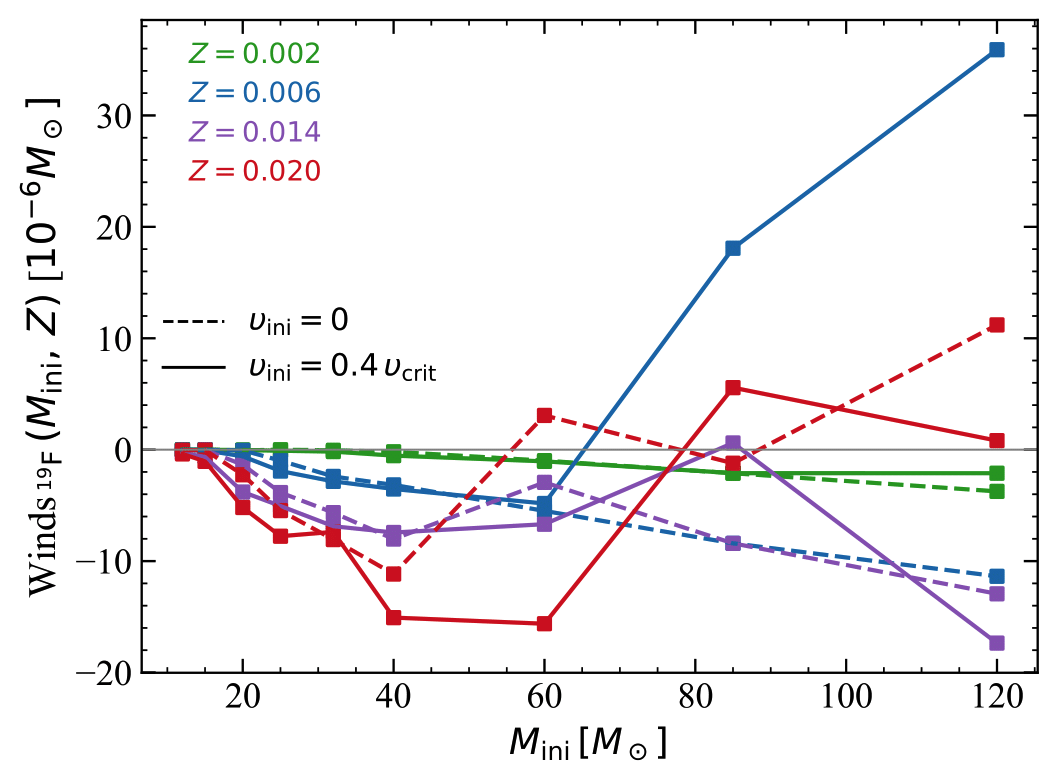}
    \caption{Wind stellar yields of ${}^{19}$F in units of $10^{-6}$~{\msol} for non-rotating (dashed curves) and moderately rotating models (solid curves) at various metallicities as a function of the initial mass of the models.}
   
    \label{fig:PIMF19W} 
    \end{figure}

    \begin{figure}
    \includegraphics[width=0.5\textwidth]{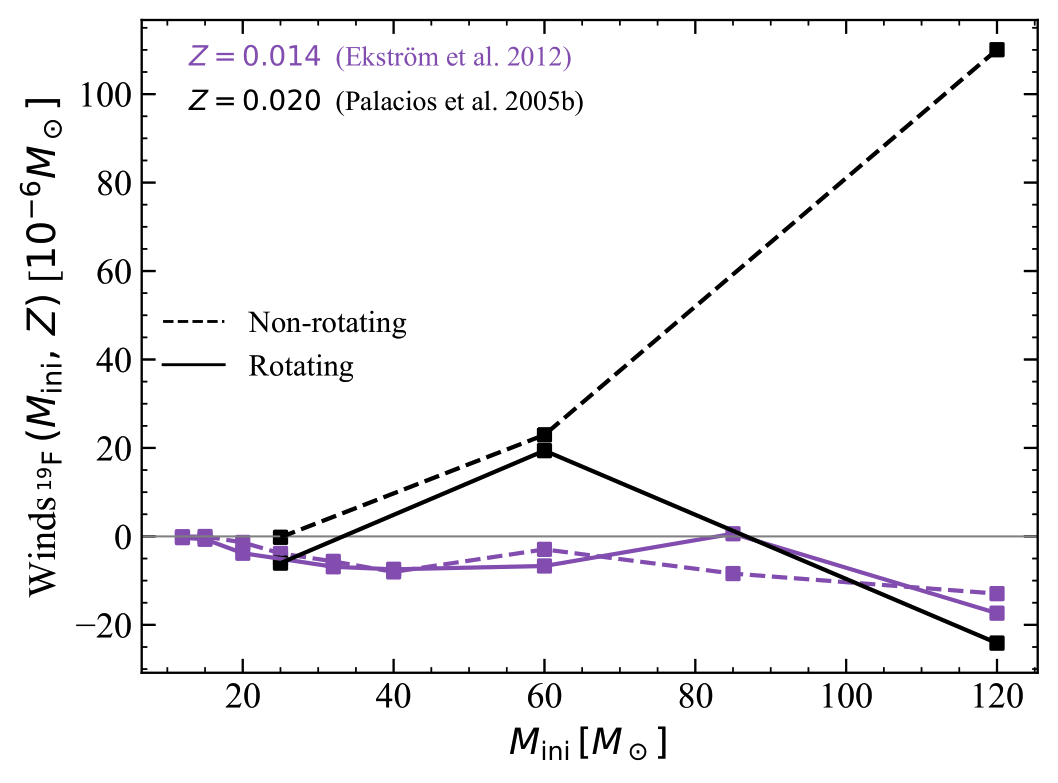}
    \caption{Comparisons between the wind stellar yields of ${}^{19}$F in units of $10^{-6}$~{\msol} obtained by this work (purple curves) from models by \citet{Ekstrom2012} and by \citet{Palacios2005} (black curves) with the same stellar evolution code for solar metallicities.}
    \label{fig:PIMF19C} 
    \end{figure}

\citet{GMMA2000} calculated yields for stellar models with initial masses ranging from 25 to 120~{\msol} and metallicities of $Z=0.008$, $0.020$ and $0.040$. Based on the mass-loss rates they adopted (see Table~\ref{table:models}), they concluded that WR stars might be significant contributors of fluorine in the present-day Universe. These yields were subsequently revised downwards by \citet{Pala2005}, who computed models with significantly lower mass-loss rates. These models also explored the effects of rotation and demonstrated that rotation plays a much smaller role than mass loss in the fluorine enrichment in the domain of metallicity (see Table~\ref{table:models}) explored in this work. 

In Table~\ref{table:yieldswindsF19}, the masses of newly synthesised ${}^{19}$F, in units of $10^{-6}$~{\msol}, ejected by stellar winds are given for various stellar models. 
For most models, the yields are negative (with positive values highlighted in boldface in Table~\ref{table:yieldswindsF19}). Positive values are rare and limited to a few models with $M_{\rm ini} \geq 60$~{\msol} and metallicities $Z \ge 0.006$. It is only in these few cases that layers enriched in fluorine appear at the stellar surface during periods of strong enough mass loss to compensate for all the phases during which depleted fluorine material is wind ejected.

Figure~\ref{fig:PIMF19W} presents these stellar yields as a function of the initial mass for different initial rotations and metallicities. Interestingly the yields for the $Z=0.006$ models for the 80~{\msol} models are larger than the yields for the same initial model at higher metallicities. This illustrates the fact that the yields of fluorine depends in an intricate way on the metallicity. Let us remind here that positive wind ${}^{19}$F yields are expected when the He-burning core is exposed at the surface at an early phase of the core He-burning phase, and when at this evolutionary stage, the stellar winds are strong. 

Figure~\ref{fig:PIMF19C} compares the wind stellar yields from this work for the solar metallicity models by \citet{Ekstrom2012} with predictions from \citet{Palacios2005}, who used {\genec} with different prescriptions for the solar composition ($Z_\odot=0.020$ instead of 0.014 in the present models) and the diffusion coefficients for the vertical shear turbulence (see Table~\ref{table:models}). We see that most of the wind stellar yields are comparable. There is one exception in the case of the 120~{\msol}. Small differences in the input physics between the two series of models has here a strong impact. Typically between these two series of models the criterion for considering a star enters the WR phase is slightly different. In the models by \citet{MM2003} used in the paper by \citet{Palacios2005}, the star enters the WR phase when it is hot ($\log{\rm T}_{\rm eff} > 4.0$ and the mass fraction of hydrogen at the surface is below 0.4. In the models of \citet{Ekstrom2012} analysed here, the star is assumed to enter in the WR phase when $\log{\rm T}_{\rm eff} > 4.0$ \citep[as in][]{MM2003} but with a mass fraction of hydrogen at the surface of 0.3. As a consequence the present models enters into the WR phase a bit later and this has here as an impact to change the evolution of the total mass of the star. The 120~{\msol} analysed in \citet{Palacios2005} ends the MS phase with a mass equal to 42.7~{\msol} and the core He-burning phase with a mass of 16.3~{\msol}, significantly smaller that the corresponding masses in the models analysed here (63.6 and 30~{\msol}, respectively). This is the main reason for the strong differences in the fluorine yields shown in Fig.~\ref{fig:PIMF19C}. For the other models the impact of the above change are much more modest. 

Since the present models predict lower yields than those by \citet{Palacios2005}, this reinforces the conclusion of these authors that winds of WRs are not significant contributors to galactic fluorine. Since the mass-loss rates of massive stars may be revised in the future, this conclusion should be viewed with some caution and be susceptible of revision. 

As mentioned in Sect.~\ref{sec:2}, \citet{Martinet2023} developed models for VMSs up to 300~{\msol} at different metallicities. These stars evolve nearly homogeneously during the MS phase due to their very large convective cores. In Table~\ref{table:yieldswindsF19} we present fluorine wind yields from these models for the metallicities $Z=0.006$ and $Z=0.014$. A striking case is the one of the 300~{\msol} model at $Z=0.006$, which will eject $7.4\times10^{-4}$~{\msol} of newly produced fluorine. But such stars are very rare and actually will not change the conclusion that winds computed with the present mass-loss rates prescription are not important sources of new fluorine at near solar metallicities. 
As already mentioned above, since the mass-loss rates are still being revised \citep[see e.g.][]{Sander2020, Sander2023}, likely such a work will be needed to be redone in the future to see whether this conclusion still holds.

\section{Fluorine content in pre-SN progenitors}\label{sec:5}

    \begin{figure}
    \includegraphics[width=0.5\textwidth]{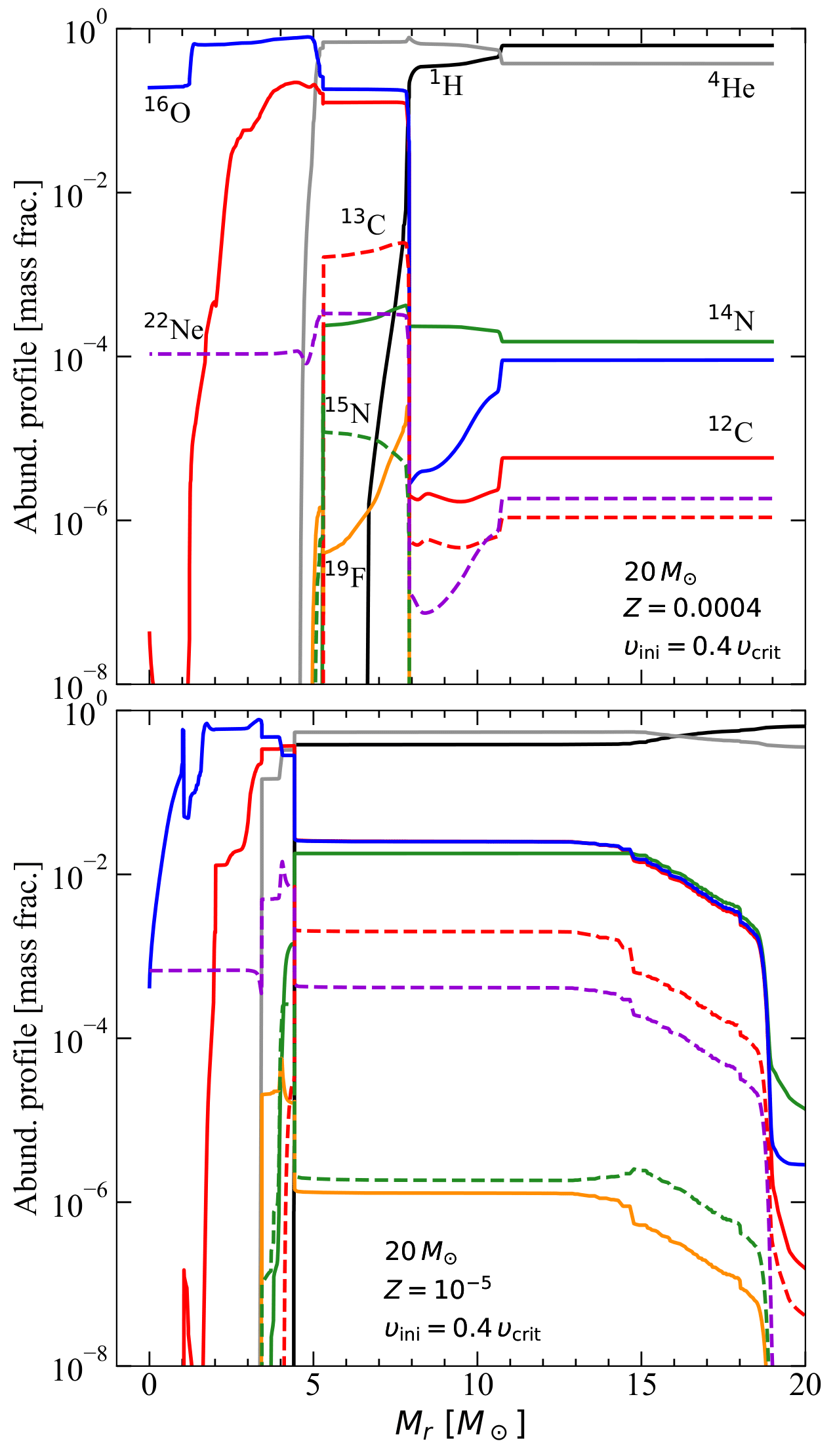} 
    \caption{Abundance profiles in the interior of 20~{\msol} moderately rotating models at IZw18 galaxy and EMP metallicities at the last computed point, during the core O-burning and the end of core O-burning phase, respectively. The isotopes shown are the same as in Fig.~\ref{fig:120Mabund}.}
    \label{fig:PIMF19W_2}
    \end{figure} 

Figure~\ref{fig:PIMF19W_2} shows the chemical abundances of metal-poor moderately rotating 20~{\msol} models during, and at the end of core O-burning phase at two different metallicities. In the model with $Z=0.0004$ metallicity (upper panel), a region enriched in fluorine (orange curve) appears in layers where helium is partially transformed into carbon (between mass coordinates 5 to 8~{\msol}). This region shows a chemical composition characteristic of an early core He-burning phase. At this metallicity and initial rotation, present models do not predict strong rotational mixing between the H- and He-burning zones, as explained below. Such mixing, when it occurs, allows for the production of significantly more nitrogen, as carbon and oxygen produced in the He-burning zone can diffuse into the H-burning shell and be transformed into nitrogen. However, in this model, the abundance of nitrogen in the matter processed by the H-burning shell is too low for such a mixing to have occurred in an important way. The level of abundance of nitrogen (green curve) in the region below the mass coordinate at 8~{\msol} is around $0.0003$ (we note that below the mass coordinate 5~{\msol}, nitrogen is completely destroyed by He-burning). Such a level of abundance can be easily explained by the transformation of the initial abundances of carbon and oxygen\footnote{If we consider that the CNO elements represents initially 3/4 of the initial total mass fraction of heavy elements (here $Z=0.0004$), we get an initial value of 0.0003 for the initial abundances of CNO. If that material is transformed into nitrogen by the CNO process, abundances of 0.0003 of nitrogen is obtained.}. Thus, in this model the mixing was not very efficient. The fluorine present between mass coordinates 5 and 8~{\msol} originates from the transformation of a part of the nitrogen. In this region, the fluorine is not destroyed by $\alpha$-captures due to insufficient temperatures (below $\sim100$ million degrees). 

The stellar yield of fluorine in this model is sensitive to the mass of the remnant. Applying the method proposed by \citet{Patton2020} \citep[as explained in][]{Sibony2023, DYS2024}, a remnant mass of 8.84~{\msol} is predicted. Conversely, using the rule suggested in \citet{Maeder1992_baryonic}, a significantly lower remnant mass is predicted of 2.13~{\msol} is predicted {\citep[more in line with the models by ][]{Limongi2018} (see Fig.~\ref{fig:Remnantmass}). Any remnant mass above 5~{\msol} significantly impacts the fluorine yields, as part of the fluorine-enriched region would remain within the remnant. A remnant mass below this threshold would minimally affect the stellar yields, allowing the entire enriched region to be ejected. This explains why, with a remnant mass of 8.84~{\msol}, the fluorine stellar yield is negative, at $-0.094\times10^{-6}$~{\msol}, while with a remnant mass of 2.13~{\msol}, the stellar yield is positive, at $9.3\times10^{-6}$~{\msol}. In this work, we used the more recent prescription by \citet{Patton2020} for computing the remnant mass. The present remark shows that such a choice has in some cases a strong impact on the fluorine yields.

The connection between the structure of a massive star at the end of its evolution and the outcome of the core collapse is the object of active researches with still many open questions. For instance, will all the mass be swallowed by the black hole, or shall we have an explosion and the formation of a neutron star, or an intermediate situation where a black hole is formed but still some layers are ejected associated with a faint SN event? Depending which one of these outcomes is realised, this may change radically the associated stellar yields, going from zero, when all the mass is retained, to full ejection in case the whole star explodes (this is expected to occur in pair-instability SNe), to intermediate situation where some mass is ejected, the part above the remnant mass. Unfortunately, these different scenarios depend on many still uncertain features of the pre-core collapse models, for instance on the entropy profile \citep[e.g.][]{Wang2022, Boccioli2023}, the distribution of the angular momentum and/or the magnetic field \citep[see e.g.][]{Ober2014, Ober2020, Griffiths2022}, the way convection is treated \citep[see e.g.][]{Rizzuti2024}, the initial metallicity \citep[][]{Chen2017}, in particular, low-metallicity rotating massive stars often develop more compact pre-SN structures that may inhibit successful explosions unless sufficiently efficient O-burning shells develop before collapse. Thus, at the moment, we are left with some very schematic guidelines linking the pre-SN structure to the final outcome. At best, they allow for some indications on how some properties of the models may impact the yields.

Rotational mixing occurs in the model shown in the lower panel of Fig.~\ref{fig:PIMF19W_2} where the metallicity is $Z=10^{-5}$. This indicates that in CNO processed regions, the maximum level of nitrogen abundance produced by the transformation of the initial carbon and oxygen abundances would be at most $7.5\times10^{-6}$ in mass fraction. However, in this model, nitrogen levels exceed this value above the mass coordinate of 3.5~{\msol}, resulting in a significant increase in nitrogen production due to rotational mixing. This process also contributes to an increase in fluorine abundance. Like the model shown in the upper panel, the stellar yields depend on the mass of the remnant. Using the methods proposed by \citet{Patton2020} and \citet{Maeder1992_baryonic}, the remnant masses are 6.98~{\msol} and 1.77~{\msol} respectively, corresponding to fluorine stellar yields of $10.58\times10^{-6}$ and $37.1\times10^{-6}$~{\msol}.

The increase observed when the metallicity decreases from $Z=0.0004$ to $Z=10^{-5}$ is mainly due to the fact that lower metallicity makes the star more compact. This change results in a smaller distance between the H- and He-burning regions in the model with $Z=10^{-5}$ compared to the model with $Z=0.0004$. The proximity facilitates mixing between these regions \citep[see a detailed discussion in][]{Tsiatsiou2024}.

Interestingly, even in the moderately rotating Pop~III 20~{\msol} model this process appears to be very inefficient. Indeed, the stellar yield for the Pop~III model is $1.7 \times 10^{-12}$~{\msol}\footnote{In Table~\ref{table:yields}, values less than $5\times10^{-9}$~{\msol} are set to 0.}. As metallicity decreases from $Z=10^{-5}$ to $Z=0$, the star becomes even more compact, which should theoretically favour mixing and, consequently, primary nitrogen production. However, this is not observed due to another factor: the efficiency of the mixing process between the H- and He-burning zones, which is notably reduced in Pop~III stars. It is important to note that mixing can be driven by convection as well as by shear turbulence. In the latter case, in the current models, the degree of differential rotation governs the efficiency of the mixing. A higher degree of differential rotation results in stronger rotational mixing. In Pop~III stars, the differential rotation is less pronounced than in models with even slight metal enrichment. In the absence of any CNO elements, core H-burning occurs at higher temperatures, sufficient to activate the triple $\alpha$ reactions, albeit at a low level. Consequently, at the end of the core H-burning phase, the core contracts only slightly before the full activation of He-burning reactions. This small contraction produces a small degree of differential rotation. Moreover, due to the low opacity of the envelope in Pop~III stars, little of the energy released by contraction is used to inflate it, which further reduces the differential rotation and thus the efficiency of the rotational mixing. A more detailed discussion is provided in \citet{Tsiatsiou2024}.

\subsection{Rapidly rotating massive Pop~III stars} \label{sec:51}

    \begin{figure}
    \includegraphics[width=0.5\textwidth]{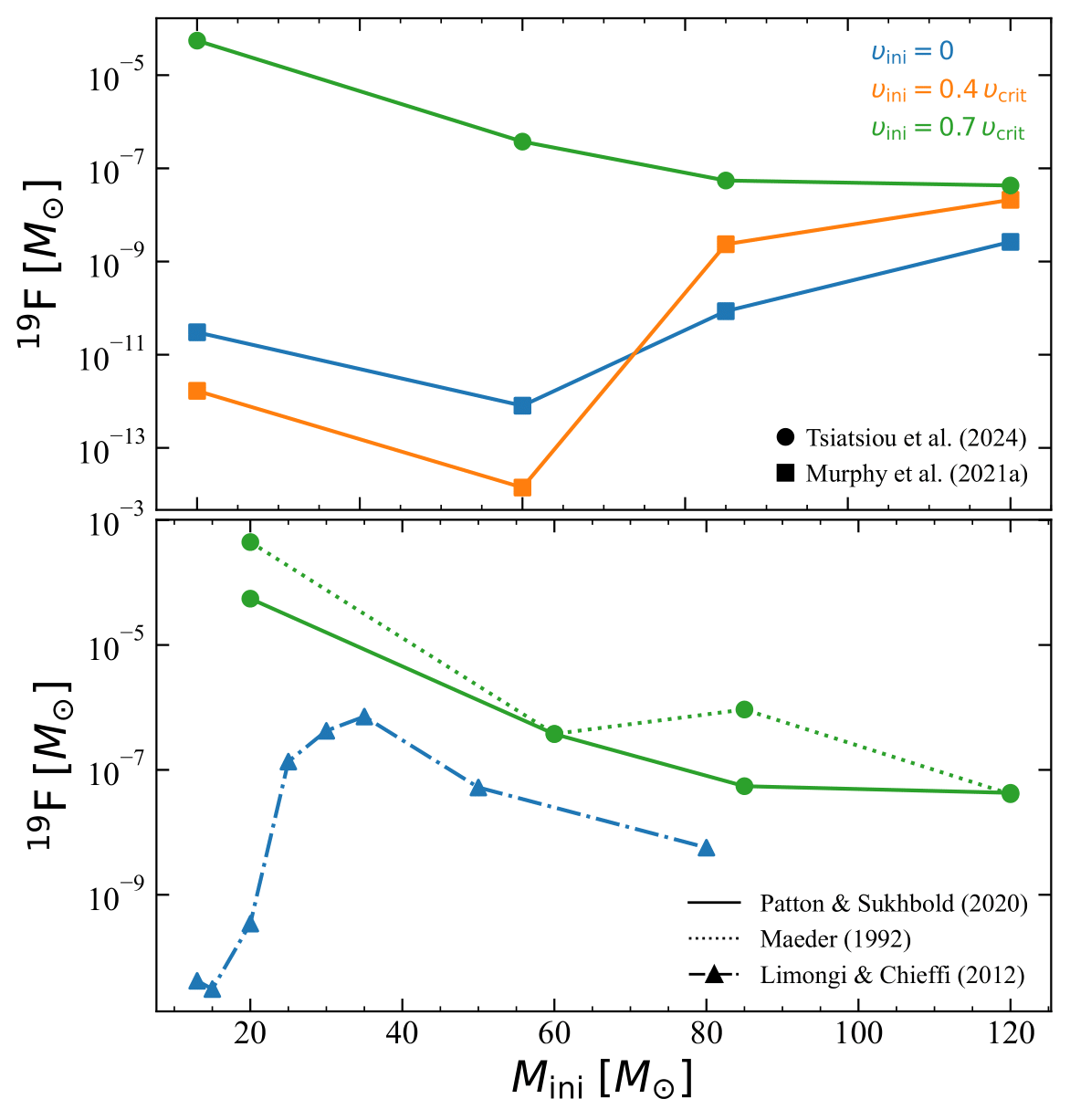}
    \caption{Stellar yields of fluorine of Pop~III models. Upper panel: Models at fast (green curves), moderate (orange curves), and zero (blue curves) rotation, with the stellar yields being calculated from \citet{Patton2020}. Bottom panel: Models at fast and zero rotation. The solid curve is from \citet{Patton2020}, dotted curve is from \citet{Maeder1992_baryonic}, and dashed-dotted curve is by \citet{Limongi2012}.}
    \label{fig:fastgrid}
    \end{figure}


\citet{Tsiatsiou2024} presents a grid of models for rapidly rotating Pop~III stellar models. Unlike the models in Table~\ref{table:yields}, these models assume an initial rotation at 70\% of the critical velocity. The fluorine yields for these rapidly rotating Pop~III stellar models are shown in Fig.~\ref{fig:fastgrid}. The upper panel compares the stellar yields from Pop~III stellar models computed with {\genec} at different initial rotations, using the method proposed by \citet{Patton2020} for determining the remnant mass. The moderately and non-rotating models were computed by \citet{Murphy2021}. We note that since we started with an initial rotation corresponding to a fixed fraction of the critical velocity on the ZAMS, this corresponds to different initial rotations depending on the initial mass. The ranges of initial rotations are indicated on the figure.

We observe that the moderately rotating models produce fluorine yields similar to those of the non-rotating ones. In contrast, there are significant enhancements in fluorine yield for the rapidly rotating models for initial masses below 80~{\msol}. This enhancement reflects the strong primary nitrogen production in rapidly rotating models, \citep[see above and][]{Tsiatsiou2024}. The yields from rapidly rotating Pop~III stellar models are comparable to those of moderately rotating models at metallicity $Z=10^{-5}$. In case Pop~III stars were indeed fast rotators, as suggested by \citet{Stacy2013a}, then the yields from the rotating $10^{-5}$ models could represent those during the early chemical evolution of Pop~III and very metal-poor massive stars. As we demonstrate below, this has significant implications for fluorine enrichment at early times.

The lower panel of Fig.~\ref{fig:fastgrid} compares the yields of the rapidly rotating Pop~III models with different assumptions for determining the mass of the remnant: the green solid curve represents the results from \citet{Patton2020} and the dotted one from \citet{Maeder1992_baryonic}. We also compare these yields with those published by \citet{Limongi2012} for non-rotating Pop~III stellar models. Clearly, lower yields are obtained when using the \citet{Patton2020} method to determine the remnant mass compared to the yields obtained with the \citet{Maeder1992_baryonic} method. As can be seen in Fig.~\ref{fig:Remnantmass}, the remnant masses given by \citet{Patton2020} are larger, thus retaining some fluorine into the remnant. The comparison with the yields from \citet{Limongi2012} shows that although their models do not account for rotation, they still obtain substantial fluorine yields. We note that the yields from \citet{Limongi2012} include contributions from explosive nucleosynthesis but do not account for any ``neutrino-induced'' reactions. These models without neutrino induced nucleosynthesis show that the explosive nucleosynthesis does not contribute to $^{19}$F.

\subsection{Very massive stars at very low metallicity}\label{sec:52}

For $Z=10^{-5}$ metallicity, rotating models are available for three initial masses above 120~{\msol} \citep{Martinet2023}: 180, 250, and 300~{\msol}. The fluorine yields for these initial masses are $10^{-9}$, $8.8\times10^{-10}$ and $1.7\times10^{-10}$~{\msol}, respectively. Comparing these with yields from lower initial masses, it is evident that these VMSs do not contribute significantly more fluorine. Using the Kroupa IMF \citep{Kroupa2001}, the number of stars with initial masses between 120 and 300~{\msol} constitutes 2.5\% of the number of stars between 9 and 120~{\msol}. Therefore, to be considered significant producers, these stars would need fluorine yields at least an order of magnitude larger than those predicted for stars with lower initial masses. However, this is not the case. 

\section{Stellar yields} \label{sec:6}

    \begin{table}
    \small
    \caption{Values of SN stellar yields of ${}^{19}{\rm F}$ in units of $10^{-6}$~{\msol} for the rapidly rotating Pop~III models.}
    \begin{center}
    \begin{tabular}{c|c}
    \hline															
    \hline \noalign{\smallskip}							
    $M_{\rm ini}$ [{\msol}]	&	$Z=0$	\\
    \hline \noalign{\smallskip}												
    \multicolumn{2}{c}{$\upsilon_{\rm ini}=0.7\,\upsilon_{\rm crit}$}	\\
    \hline \noalign{\smallskip}
    9 &   0.0  \\ [0.4ex]
    20 &  55.21  \\ [0.4ex]
    60 &  0.374  \\ [0.4ex]
    85 &  0.055  \\ [0.4ex]
    120 & 	0.043 \\ [0.4ex]
    \hline
    \end{tabular}
    \end{center}
    \label{table:yieldsPopIII}
    \end{table}

Table~\ref{table:yields} presents the SN stellar yields for our models. Looking at Table~\ref{table:yields}, we observe the following features:
\begin{itemize}
    \item Models with metallicities $Z<0.002$ produce positive stellar yields in fluorine.
    \item The contribution of models at $Z>0.002$ with masses below 25~{\msol} are either negative or modest. However, above 25~{\msol}, a few models, such as the rotating 32 and 40~{\msol}, show significant stellar yields.
    \item Lower initial mass models (favoured by the IMF) start to show positive fluorine stellar yields at $Z \leq 0.0004$.
    \item The metallicity $Z=10^{-5}$ is the most favourable in our current grids for the production of fluorine, particularly when rotational mixing is considered. At this metallicity, the H- and He-burning zones are separated by a small radiative layer where the shear turbulence is strong enough to drive mixing. At higher metalliticies, the increased distance discourages mixing, while at lower metallicities, decreased shear turbulence also hampers mixing.
    \item Pop~III models, due to the reason explained above, do not exhibit strong fluorine production. However, this is dependent upon the initial rotation adopted. Considering higher initial rotation alters this outcome.
\end{itemize}

In order to estimate the impacts of spinstars on fluorine abundance, we follow a very simple chemical evolution model that assumes instantaneous recycling and mixing. According to \citet{Tinsley1980}, the mass fraction of a given element in the ISM at any given time (t) can be expressed as
\begin{equation}
    X_{19}(t)=p_{19}^{\rm net}\ln{\left (\frac{1}{\sigma(t)} \right )},
\end{equation}
where $\sigma(t)$ is the mass fraction of the gas in the Galaxy at time $t$, and $p_{19}^{\rm net}$ represents the net stellar yields, calculated by:
\begin{dmath} \label{eq:4}
    p_{19}^{net}(Z)= \frac{1}{1-R} \int_{9}^{120} \left (p_{19}^{\rm SN}(M_{\rm ini},Z)+p_{19}^{\rm winds}(M_{\rm ini},Z) \right ) \, \phi(M_{\rm ini}) \, {\rm d}M_{\rm ini}
\end{dmath} 
Here, $R$ is the returned mass fraction and $\phi$ the initial mass function. The returned mass fraction is defined by
\begin{equation}
    R=\int_{9}^{120} (M_{\rm ini}-M_{\rm cut}) \, \phi (M_{\rm ini}) \, {\rm d}M_{\rm ini},
\end{equation}
which represents the total mass ejected by a stellar population divided by the total mass used to form the stars. It is important to note that the integral of $\int_{0.01}^{120} \, \phi (M_{\rm ini}) \, {\rm d} M_{\rm ini}$ is imposed to be equal to 1. As we are focusing on the early phases of chemical enrichment, we only consider the mass ejected by the short-lived massive stars, assuming the mass locked into low initial mass stars remains in the form of stars.

Although $p_{19}^{\rm net}$ should ideally be a time-independant quantity, this is not the case in practice. We assume that the yields for $Z=10^{-5}$ metallicity are representative for an initial period of enrichment by SN ejecta. According to \citet{Prantzos1995}, the ratio of the mass of gas over to the total mass of gas and stars is approximately 0.2. 
This ratio was larger in the early phase of the galaxy. 
Let us take a value equal to 0.9, 
following \citet{Prantzos1995}. Consequently, we obtain $X_{19}(t)=5.8\times10^{-7}$, a mass fraction similar to that of fluorine in the solar neighbourhood. Thus, this simple estimate provides some confidence that fluorine synthesised by rotating very metal-poor stars may indeed have significantly contributed to the abundance of fluorine in the Universe. Of course, more sophisticated chemical evolution models of the Milky Way are needed to really assess the impact of the present fluorine yields as those of \citet[see e.g][]{Chiapp2015, Prantzos2023, Koba2023, Fernandez2024}.

    \begin{figure}
    \includegraphics[width=0.45\textwidth]{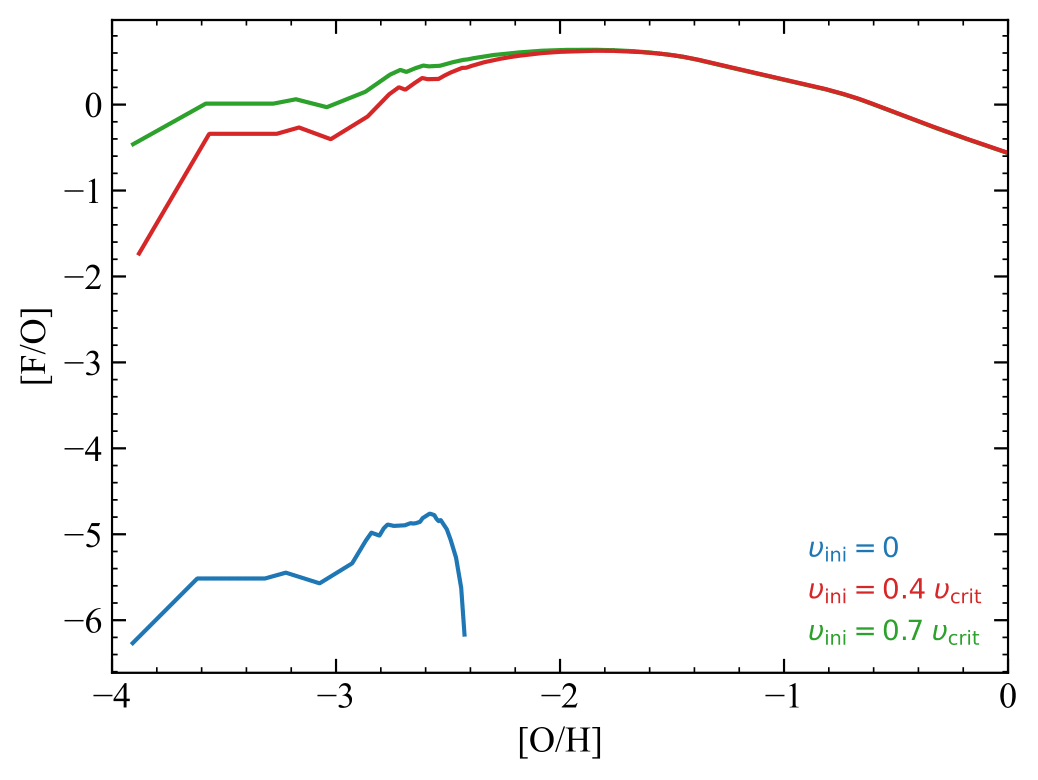}
    \caption{Evolution of [F/O] as a function of [O/H] from stellar yields in the metallicity range of $0 \leq Z \leq 0.020$ with different initial rotations.}
    \label{fig:gce}
    \end{figure}

Figure~\ref{fig:gce} shows [F/O] as a function of [O/H] computed in one-zone galactic chemical evolution (GCE) model in the chemical evolution library \citep[\textsc{celib},][]{Saitoh2017, Hirai2021} \citep[as in][]{Tsiatsiou2024}, using yields from models with different rotations. We assume that all the massive stars contribute to the chemical evolution. We note that no weighting is applied to different initial rotation velocities. Furthermore, we assume that all stars explode as SNe, with no separate treatment of stellar winds for the most massive objects. This approach is deliberately schematic, intended to illustrate the significant impact of different rotation velocities on the stellar yields. More sophisticated GCE models could adopt a realistic distribution of rotation rates and more refined assumptions about mass loss and final fates. The green curve corresponds to the case where for metallicities $Z < 10^{-5}$ the yields of the rapidly rotating Pop~III stellar models are used. For metallicities $Z \geq 10^{-5}$, we adopted the yields of the moderately rotating models. 

As shown in Fig.~\ref{fig:gce}, [F/O] ratios are significantly higher in models accounting for SN stellar yields coming from rotating stellar models. Moreover, for those models, a similar behaviour is observed in the evolution of fluorine as metallicity increases, analogous to the trend seen for nitrogen in \citet{Tsiatsiou2024}. The rapidly rotating models show an enhancement of [F/O] ratios for [O/H] $\leq -1.9$. In contrast, the non-rotating models with higher metallicities ($Z \geq 0.0004$) destroy the fluorine abundances, preventing them from contributing to the GCE model (see Table~\ref{table:yields}). [F/O] ratios in the moderately rotating models peak at [O/H] = $-2$, where mixing between H- and He-burning zones is the most efficient \citep{Tsiatsiou2024}. As we apply the same yields for stars with $Z \geq 10^{-5}$, [F/O] ratios in models with rotation follow a consistent trend for [O/H] $\geq -2$.

These results support previous studies that highlight the necessity of rotating massive stars at low metallicity \citep[e.g.][]{Prantzos2018, Ryde2020, Grisoni2020, Womack2023, BijavaraSeshashayana2024}. \citet{Grisoni2020} concluded that rotating massive stars significantly contribute to the enrichment of fluorine in environments with low metallicity using their GCE models. \citet{Womack2023}, using different combinations of yields for AGB and massive stars, conclude that the observed trends of the ratios [F/O], [F/Fe] abundance ratios at low metallicity, and the increase observed for [F/Ba] at [Fe/H] $\geq -1$ favour models with contributions from rapidly rotating massive stars at all metallicities.\citet{BijavaraSeshashayana2024} demonstrated that the production of fluorine in AGB stars \citep[e.g.][]{Karakas2018} and rotating massive stars \citep{Limongi2018} could reasonably explain the observed fluorine abundances in open clusters.

Our study specifically focuses on the behaviour of [F/O] ratios predicted from our yields. The GCE models considering the inhomogeneity of the ISM \citep[e.g.][]{Cescutti2013, Hirai2019} is required to allow for a more reliable comparison with observational data. Additionally, given our prediction of an enhanced and slightly decreasing trend of [F/O] ratios at higher metallicity, it is of prime interest to obtain the near-infrared high-dispersion spectroscopic measurements of fluorine in metal-poor stars to better understand the origin of fluorine.

\section{Discussion and conclusions} \label{sec:7}

We have discussed the synthesis of fluorine in massive star models with and without rotation and at different initial metallicities. The main results are the following:
\begin{enumerate}
    \item The surface abundances of fluorine during the MS phase, the RSG phase, or the early WC phase of massive stars provide critical insights into the transformation of fluorine in the H- and He-burning zones. 
    \item Winds appear to be inefficient at enriching the stellar surroundings with newly synthesised fluorine. Previous studies that highlighted WR winds as a significant source at solar metallicity relied on stronger mass-loss rates that current prescriptions in stellar models no longer support. While rotation assists in ejecting fluorine, it does not fully compensate for the reduced impact of stellar winds. Consequently, even rotating models are unable to make a significant contribution.
    \item We have presented the results for VMSs with masses up to 300~{\msol}. Although some models may show large fluorine yields, the rarity of such stars means they are less significant galactic contributors compared to the more numerous lower-mass stars.
    \item We find that the fluorine yields coming from the SN ejecta of very metal-poor rotating massive stars are sufficiently high to significantly contribute to the enrichment of the ISM. It is mainly produced in regions in between the He- and H-burning shell. We emphasise, however, that these yields remain sensitive to the assumptions made about the stellar fate (e.g. mass-cut in SN models and the possibility that some stars fail to explode and instead undergo direct collapse). It would be interesting to incorporate these yields into a more sophisticated GCE model in order to determine whether these yields significantly impact the observed fluorine levels or to predict fluorine levels in metal-poor halo stars.    
    \item Our findings underscore the complementary roles of AGB stars and massive stars in fluorine production across different metallicities. While AGB stars are established as significant contributors to fluorine synthesis, particularly at near-solar metallicities, our study reveals that massive stars can notably contribute at lower metallicities under certain high mass and fast rotation conditions. Specifically, the yields from SNe from rapidly rotating massive star progenitors enhance fluorine abundance under these conditions, suggesting a broader production landscape with massive stars potentially filling a role in environments where AGB stars may be less effective.
\end{enumerate}

Many uncertainties influence the conclusions drawn above due to subtleties in the nuclear fluorine production process that are not usually accounted for in stellar modelling, including the crucial necessity of very precisely predicting the boundaries between H- and He-rich zones. The very special role played by wind mass losses in fluorine survival is an additional important difficulty. Many multidimensional effects that are not considered in our models may consequently have a strong impact on the fluorine yields. Namely, they concern the turbulent transport of nuclides in different stellar zones, the boundaries of which are very uncertain in our models; the proper treatment of rotation; or even the role of magnetic fields. These multidimensional effects may make the traditional SN onion-skin structure completely obsolete. Wind mass losses are probably much more complex than considered here. Clumpiness or the influence of binarity may drastically affect the wind predictions \citep[see e.g.][]{Brinkman2023}, and binary interactions can also produce appreciable amounts of fluorine under certain conditions \citep[see Sect.~4.2 in.][]{Brinkman2023}. Finally, the predicted role of the different stars considered here based on the adopted highly simplistic model for the evolution of the chemical content of the Galaxy is by far not very reliable. In particular, many effects related to the dynamics of the Galaxy should not have to be omitted. Overall, the changes in nuclear reaction rates involving ${}^{19}{\rm F}$ published since NACRE (\citet{O18pg19F2012, F19palphaO162017, F19palphaO162018, Williams2021_rates, F19palphaO162021} for H-burning regions and \citet{Ugalde2008, Iliadis2010} for He-burning regions) do not significantly impact the conclusions derived in this study.

A topic not addressed in this work but slated for future exploration is the impact of these results on explaining fluorine abundances in CEMP stars. To date, fluorine analysis in CEMP stars is limited. However, a few cases exist \citep{Frebel2015}, and these will be discussed in a forthcoming paper dedicated to CEMP stars.


\begin{acknowledgements}
We are grateful to the referee, Dr. Lorenzo Roberti for useful comments and suggestions, which have improved the manuscript. ST, GM, SE, and YS have received funding from the European Research Council (ERC) under the European Union's Horizon 2020 research and innovation program (grant agreement No 833925, project STAREX). SE has received support from the SNF project No 212143. YH has been supported by JSPS KAKENHI Grant Numbers JP22KJ0157, JP21H04499, JP21K03614, and JP22H01259. AC is post-doctorate F.R.S-FNRS fellow. RG gratefully acknowledges the grants support provided by FAPERJ under the PDR-10 grant number E26-205.964/2022 and ANID Fondecyt Postdoc No. 3230001. J.G.F-T gratefully acknowledges the grant support provided by Proyecto Fondecyt Iniciaci\'on No. 11220340, and from the Joint Committee ESO-Government of Chile 2021 (ORP 023/2021) and 2023 (ORP 062/2023). EF is support by SNF grant number 200020\_212124.
\end{acknowledgements}


\bibliographystyle{aa_url} 
\bibliography{arXiv}

\begin{thebibliography}{119}
\expandafter\ifx\csname natexlab\endcsname\relax\def\natexlab#1{#1}\fi

\bibitem[{{Abia} {et~al.}(2019){Abia}, {Cristallo}, {Cunha}, {de Laverny}, \&
  {Smith}}]{Abia2019}
{Abia}, C., {Cristallo}, S., {Cunha}, K., {de Laverny}, P., \& {Smith}, V.~V.
  2019,
  \href{http://dx.doi.org/10.1051/0004-6361/201935286}{\color{magenta}\aap},
  \href{https://ui.adsabs.harvard.edu/abs/2019A&A...625A..40A}{625, A40}

\bibitem[{{Abia} {et~al.}(2015{\natexlab{a}}){Abia}, {Cunha}, {Cristallo}, \&
  {de Laverny}}]{Abia2015}
{Abia}, C., {Cunha}, K., {Cristallo}, S., \& {de Laverny}, P.
  2015{\natexlab{a}},
  \href{http://dx.doi.org/10.1051/0004-6361/201526586}{\color{magenta}\aap},
  \href{https://ui.adsabs.harvard.edu/abs/2015A&A...581A..88A}{581, A88}

\bibitem[{{Abia} {et~al.}(2015{\natexlab{b}}){Abia}, {Cunha}, {Cristallo}, \&
  {de Laverny}}]{AbiaCORR2015}
{Abia}, C., {Cunha}, K., {Cristallo}, S., \& {de Laverny}, P.
  2015{\natexlab{b}},
  \href{http://dx.doi.org/10.1051/0004-6361/201526586e}{\color{magenta}\aap},
  \href{https://ui.adsabs.harvard.edu/abs/2015A&A...584C...1A}{584, C1}

\bibitem[{{Abia} {et~al.}(2010){Abia}, {Cunha}, {Cristallo}, {de Laverny},
  {Dom{\'\i}nguez}, {Eriksson}, {Gialanella}, {Hinkle}, {Imbriani},
  {Recio-Blanco}, {Smith}, {Straniero}, \& {Wahlin}}]{Abia2010}
{Abia}, C., {Cunha}, K., {Cristallo}, S., {et~al.} 2010,
  \href{http://dx.doi.org/10.1088/2041-8205/715/2/L94}{\color{magenta}\apjl},
  \href{https://ui.adsabs.harvard.edu/abs/2010ApJ...715L..94A}{715, L94}

\bibitem[{{Abia} {et~al.}(2009){Abia}, {Recio-Blanco}, {de Laverny},
  {Cristallo}, {Dom{\'\i}nguez}, \& {Straniero}}]{Abia2009}
{Abia}, C., {Recio-Blanco}, A., {de Laverny}, P., {et~al.} 2009,
  \href{http://dx.doi.org/10.1088/0004-637X/694/2/971}{\color{magenta}\apj},
  \href{https://ui.adsabs.harvard.edu/abs/2009ApJ...694..971A}{694, 971}

\bibitem[{{Alves-Brito} {et~al.}(2011){Alves-Brito}, {Karakas}, {Yong},
  {Mel{\'e}ndez}, \& {V{\'a}squez}}]{Alves2011}
{Alves-Brito}, A., {Karakas}, A.~I., {Yong}, D., {Mel{\'e}ndez}, J., \&
  {V{\'a}squez}, S. 2011,
  \href{http://dx.doi.org/10.1051/0004-6361/201116604}{\color{magenta}\aap},
  \href{https://ui.adsabs.harvard.edu/abs/2011A&A...536A..40A}{536, A40}

\bibitem[{{Alves-Brito} {et~al.}(2012){Alves-Brito}, {Yong}, {Mel{\'e}ndez},
  {V{\'a}squez}, \& {Karakas}}]{Alves2012}
{Alves-Brito}, A., {Yong}, D., {Mel{\'e}ndez}, J., {V{\'a}squez}, S., \&
  {Karakas}, A.~I. 2012,
  \href{http://dx.doi.org/10.1051/0004-6361/201118623}{\color{magenta}\aap},
  \href{https://ui.adsabs.harvard.edu/abs/2012A&A...540A...3A}{540, A3}

\bibitem[{{Angulo} {et~al.}(1999){Angulo}, {Arnould}, {Rayet}, {Descouvemont},
  {Baye}, {Leclercq-Willain}, {Coc}, {Barhoumi}, {Aguer}, {Rolfs}, {Kunz},
  {Hammer}, {Mayer}, {Paradellis}, {Kossionides}, {Chronidou}, {Spyrou},
  {degl'Innocenti}, {Fiorentini}, {Ricci}, {Zavatarelli}, {Providencia},
  {Wolters}, {Soares}, {Grama}, {Rahighi}, {Shotter}, \& {Lamehi
  Rachti}}]{Angulo1999NACRE}
{Angulo}, C., {Arnould}, M., {Rayet}, M., {et~al.} 1999,
  \href{http://dx.doi.org/10.1016/S0375-9474(99)00030-5}{\color{magenta}\nphysa},
  \href{https://ui.adsabs.harvard.edu/abs/1999NuPhA.656....3A}{656, 3}

\bibitem[{{Battino} {et~al.}(2022){Battino}, {Pignatari}, {Tattersall},
  {Denissenkov}, \& {Herwig}}]{Battino2022}
{Battino}, U., {Pignatari}, M., {Tattersall}, A., {Denissenkov}, P., \&
  {Herwig}, F. 2022,
  \href{http://dx.doi.org/10.3390/universe8030170}{\color{magenta}Universe},
  \href{https://ui.adsabs.harvard.edu/abs/2022Univ....8..170B}{8, 170}

\bibitem[{{Battino} {et~al.}(2019){Battino}, {Tattersall}, {Lederer-Woods},
  {Herwig}, {Denissenkov}, {Hirschi}, {Trappitsch}, {den Hartogh}, {Pignatari},
  \& {NuGrid Collaboration}}]{Battinoup2019}
{Battino}, U., {Tattersall}, A., {Lederer-Woods}, C., {et~al.} 2019,
  \href{http://dx.doi.org/10.1093/mnras/stz2158}{\color{magenta}\mnras},
  \href{https://ui.adsabs.harvard.edu/abs/2019MNRAS.489.1082B}{489, 1082}

\bibitem[{{Bijavara Seshashayana} {et~al.}(2024){Bijavara Seshashayana},
  {J{\"o}nsson}, {D'Orazi}, {Nandakumar}, {Oliva}, {Bragaglia}, {Sanna},
  {Romano}, {Spitoni}, {Karakas}, {Lugaro}, \&
  {Origlia}}]{BijavaraSeshashayana2024}
{Bijavara Seshashayana}, S., {J{\"o}nsson}, H., {D'Orazi}, V., {et~al.} 2024,
  \href{http://dx.doi.org/10.1051/0004-6361/202349068}{\color{magenta}\aap},
  \href{https://ui.adsabs.harvard.edu/abs/2024A&A...683A.218B}{683, A218}

\bibitem[{{Boccioli} {et~al.}(2023){Boccioli}, {Roberti}, {Limongi}, {Mathews},
  \& {Chieffi}}]{Boccioli2023}
{Boccioli}, L., {Roberti}, L., {Limongi}, M., {Mathews}, G.~J., \& {Chieffi},
  A. 2023,
  \href{http://dx.doi.org/10.3847/1538-4357/acc06a}{\color{magenta}\apj},
  \href{https://ui.adsabs.harvard.edu/abs/2023ApJ...949...17B}{949, 17}

\bibitem[{{Brinkman} {et~al.}(2023){Brinkman}, {Doherty}, {Pignatari}, {Pols},
  \& {Lugaro}}]{Brinkman2023}
{Brinkman}, H.~E., {Doherty}, C., {Pignatari}, M., {Pols}, O., \& {Lugaro}, M.
  2023, \href{http://dx.doi.org/10.3847/1538-4357/acd7ea}{\color{magenta}\apj},
  \href{https://ui.adsabs.harvard.edu/abs/2023ApJ...951..110B}{951, 110}

\bibitem[{{Buckner} {et~al.}(2012){Buckner}, {Iliadis}, {Cesaratto}, {Howard},
  {Clegg}, {Champagne}, \& {Daigle}}]{O18pg19F2012}
{Buckner}, M.~Q., {Iliadis}, C., {Cesaratto}, J.~M., {et~al.} 2012,
  \href{http://dx.doi.org/10.1103/PhysRevC.86.065804}{\color{magenta}\prc},
  \href{https://ui.adsabs.harvard.edu/abs/2012PhRvC..86f5804B}{86, 065804}

\bibitem[{{Cescutti} {et~al.}(2013){Cescutti}, {Chiappini}, {Hirschi},
  {Meynet}, \& {Frischknecht}}]{Cescutti2013}
{Cescutti}, G., {Chiappini}, C., {Hirschi}, R., {Meynet}, G., \&
  {Frischknecht}, U. 2013,
  \href{http://dx.doi.org/10.1051/0004-6361/201220809}{\color{magenta}\aap},
  \href{https://ui.adsabs.harvard.edu/abs/2013A&A...553A..51C}{553, A51}

\bibitem[{{Chen} {et~al.}(2017){Chen}, {Heger}, {Whalen}, {Moriya}, {Bromm}, \&
  {Woosley}}]{Chen2017}
{Chen}, K.-J., {Heger}, A., {Whalen}, D.~J., {et~al.} 2017,
  \href{http://dx.doi.org/10.1093/mnras/stx470}{\color{magenta}\mnras},
  \href{https://ui.adsabs.harvard.edu/abs/2017MNRAS.467.4731C}{467, 4731}

\bibitem[{{Chiappini} {et~al.}(2015){Chiappini}, {Anders}, {Rodrigues},
  {Miglio}, {Montalb{\'a}n}, {Mosser}, {Girardi}, {Valentini}, {Noels},
  {Morel}, {Minchev}, {Steinmetz}, {Santiago}, {Schultheis}, {Martig}, {da
  Costa}, {Maia}, {Allende Prieto}, {de Assis Peralta}, {Hekker},
  {Theme{\ss}l}, {Kallinger}, {Garc{\'\i}a}, {Mathur}, {Baudin}, {Beers},
  {Cunha}, {Harding}, {Holtzman}, {Majewski}, {M{\'e}sz{\'a}ros}, {Nidever},
  {Pan}, {Schiavon}, {Shetrone}, {Schneider}, \& {Stassun}}]{Chiapp2015}
{Chiappini}, C., {Anders}, F., {Rodrigues}, T.~S., {et~al.} 2015,
  \href{http://dx.doi.org/10.1051/0004-6361/201525865}{\color{magenta}\aap},
  \href{https://ui.adsabs.harvard.edu/abs/2015A&A...576L..12C}{576, L12}

\bibitem[{{Choplin} {et~al.}(2018){Choplin}, {Hirschi}, {Meynet},
  {Ekstr{\"o}m}, {Chiappini}, \& {Laird}}]{Choplin2018}
{Choplin}, A., {Hirschi}, R., {Meynet}, G., {et~al.} 2018,
  \href{http://dx.doi.org/10.1051/0004-6361/201833283}{\color{magenta}\aap},
  \href{https://ui.adsabs.harvard.edu/abs/2018A&A...618A.133C}{618, A133}

\bibitem[{{Conti}(1988)}]{Conti1988}
{Conti}, P.~S. 1988, in NASA Special Publication, ed. D.~{Baade}, P.~S.
  {Conti}, L.~{Divan}, C.~D. {Garmany}, H.~F. {Henrichs}, R.~P. {Kudritzki},
  A.~{Pauldrach}, M.~L. {Pr{\'e}vot-Burnichon}, J.~{Puls}, A.~B. {Underhill},
  \& R.~N. {Thomas}, Vol. 497, 168

\bibitem[{{Cristallo} {et~al.}(2014){Cristallo}, {Di Leva}, {Imbriani},
  {Piersanti}, {Abia}, {Gialanella}, \& {Straniero}}]{Cristallo2014}
{Cristallo}, S., {Di Leva}, A., {Imbriani}, G., {et~al.} 2014,
  \href{http://dx.doi.org/10.1051/0004-6361/201424370}{\color{magenta}\aap},
  \href{https://ui.adsabs.harvard.edu/abs/2014A&A...570A..46C}{570, A46}

\bibitem[{{Cunha} {et~al.}(2003){Cunha}, {Smith}, {Lambert}, \&
  {Hinkle}}]{Cunha2003}
{Cunha}, K., {Smith}, V.~V., {Lambert}, D.~L., \& {Hinkle}, K.~H. 2003,
  \href{http://dx.doi.org/10.1086/377023}{\color{magenta}\aj},
  \href{https://ui.adsabs.harvard.edu/abs/2003AJ....126.1305C}{126, 1305}

\bibitem[{{de Jager} {et~al.}(1988){de Jager}, {Nieuwenhuijzen}, \& {van der
  Hucht}}]{deJager1988}
{de Jager}, C., {Nieuwenhuijzen}, H., \& {van der Hucht}, K.~A. 1988, \aaps,
  \href{https://ui.adsabs.harvard.edu/abs/1988A&AS...72..259D}{72, 259}

\bibitem[{{de Laverny} \& {Recio-Blanco}(2013)}]{Laverny2013}
{de Laverny}, P. \& {Recio-Blanco}, A. 2013,
  \href{http://dx.doi.org/10.1051/0004-6361/201321491}{\color{magenta}\aap},
  \href{https://ui.adsabs.harvard.edu/abs/2013A&A...555A.121D}{555, A121}

\bibitem[{{Eggenberger} {et~al.}(2021){Eggenberger}, {Ekstr{\"o}m}, {Georgy},
  {Martinet}, {Pezzotti}, {Nandal}, {Meynet}, {Buldgen}, {Salmon},
  {Haemmerl{\'e}}, {Maeder}, {Hirschi}, {Yusof}, {Groh}, {Farrell}, {Murphy},
  \& {Choplin}}]{Eggenberger2021}
{Eggenberger}, P., {Ekstr{\"o}m}, S., {Georgy}, C., {et~al.} 2021,
  \href{http://dx.doi.org/10.1051/0004-6361/202141222}{\color{magenta}\aap},
  \href{https://ui.adsabs.harvard.edu/abs/2021A&A...652A.137E}{652, A137}

\bibitem[{{Eggenberger} {et~al.}(2008){Eggenberger}, {Meynet}, {Maeder},
  {Hirschi}, {Charbonnel}, {Talon}, \& {Ekstr{\"o}m}}]{Eggenberger2008}
{Eggenberger}, P., {Meynet}, G., {Maeder}, A., {et~al.} 2008,
  \href{http://dx.doi.org/10.1007/s10509-007-9511-y}{\color{magenta}\apss},
  \href{https://ui.adsabs.harvard.edu/abs/2008Ap&SS.316...43E}{316, 43}

\bibitem[{{Ekstr{\"o}m} {et~al.}(2012){Ekstr{\"o}m}, {Georgy}, {Eggenberger},
  {Meynet}, {Mowlavi}, {Wyttenbach}, {Granada}, {Decressin}, {Hirschi},
  {Frischknecht}, {Charbonnel}, \& {Maeder}}]{Ekstrom2012}
{Ekstr{\"o}m}, S., {Georgy}, C., {Eggenberger}, P., {et~al.} 2012,
  \href{http://dx.doi.org/10.1051/0004-6361/201117751}{\color{magenta}\aap},
  \href{https://ui.adsabs.harvard.edu/abs/2012A&A...537A.146E}{537, A146}

\bibitem[{{Federman} {et~al.}(2005){Federman}, {Sheffer}, {Lambert}, \&
  {Smith}}]{Federman2005}
{Federman}, S.~R., {Sheffer}, Y., {Lambert}, D.~L., \& {Smith}, V.~V. 2005,
  \href{http://dx.doi.org/10.1086/426778}{\color{magenta}\apj},
  \href{https://ui.adsabs.harvard.edu/abs/2005ApJ...619..884F}{619, 884}

\bibitem[{{Fernandes de Melo} {et~al.}(2024){Fernandes de Melo}, {Lombardo},
  {Alencastro Puls}, {Romano}, {Hansen}, {Tsiatsiou}, \&
  {Meynet}}]{Fernandez2024}
{Fernandes de Melo}, R., {Lombardo}, L., {Alencastro Puls}, A., {et~al.} 2024,
  \href{http://dx.doi.org/10.1051/0004-6361/202451173}{\color{magenta}\aap},
  \href{https://ui.adsabs.harvard.edu/abs/2024A&A...691A.220F}{691, A220}

\bibitem[{{Frebel} \& {Norris}(2015)}]{Frebel2015}
{Frebel}, A. \& {Norris}, J.~E. 2015,
  \href{http://dx.doi.org/10.1146/annurev-astro-082214-122423}{\color{magenta}\araa},
  \href{https://ui.adsabs.harvard.edu/abs/2015ARA&A..53..631F}{53, 631}

\bibitem[{{Georgy} {et~al.}(2013){Georgy}, {Ekstr{\"o}m}, {Eggenberger},
  {Meynet}, {Haemmerl{\'e}}, {Maeder}, {Granada}, {Groh}, {Hirschi}, {Mowlavi},
  {Yusof}, {Charbonnel}, {Decressin}, \& {Barblan}}]{Georgy2013a}
{Georgy}, C., {Ekstr{\"o}m}, S., {Eggenberger}, P., {et~al.} 2013, \aap, 558,
  A103

\bibitem[{{Georgy} {et~al.}(2014){Georgy}, {Saio}, \& {Meynet}}]{Georgy2014}
{Georgy}, C., {Saio}, H., \& {Meynet}, G. 2014,
  \href{http://dx.doi.org/10.1093/mnrasl/slt165}{\color{magenta}\mnras},
  \href{https://ui.adsabs.harvard.edu/abs/2014MNRAS.439L...6G}{439, L6}

\bibitem[{{Goriely} {et~al.}(1989){Goriely}, {Jorissen}, \&
  {Arnould}}]{Goriely1989}
{Goriely}, S., {Jorissen}, A., \& {Arnould}, M. 1989, in Nuclear Astrophysics,
  \href{https://ui.adsabs.harvard.edu/abs/1989nuas.conf...60G}{60}

\bibitem[{{Goriely} \& {Mowlavi}(2000)}]{Goriely2000}
{Goriely}, S. \& {Mowlavi}, N. 2000, \aap,
  \href{https://ui.adsabs.harvard.edu/abs/2000A&A...362..599G}{362, 599}

\bibitem[{{Griffiths} {et~al.}(2022){Griffiths}, {Eggenberger}, {Meynet},
  {Moyano}, \& {Aloy}}]{Griffiths2022}
{Griffiths}, A., {Eggenberger}, P., {Meynet}, G., {Moyano}, F., \& {Aloy},
  M.-{\'A}. 2022,
  \href{http://dx.doi.org/10.1051/0004-6361/202243599}{\color{magenta}\aap},
  \href{https://ui.adsabs.harvard.edu/abs/2022A&A...665A.147G}{665, A147}

\bibitem[{{Grisoni} {et~al.}(2020){Grisoni}, {Romano}, {Spitoni}, {Matteucci},
  {Ryde}, \& {J{\"o}nsson}}]{Grisoni2020}
{Grisoni}, V., {Romano}, D., {Spitoni}, E., {et~al.} 2020,
  \href{http://dx.doi.org/10.1093/mnras/staa2316}{\color{magenta}\mnras},
  \href{https://ui.adsabs.harvard.edu/abs/2020MNRAS.498.1252G}{498, 1252}

\bibitem[{{Groh} {et~al.}(2019){Groh}, {Ekstr{\"o}m}, {Georgy}, {Meynet},
  {Choplin}, {Eggenberger}, {Hirschi}, {Maeder}, {Murphy}, {Boian}, \&
  {Farrell}}]{Groh2019}
{Groh}, J.~H., {Ekstr{\"o}m}, S., {Georgy}, C., {et~al.} 2019,
  \href{http://dx.doi.org/10.1051/0004-6361/201833720}{\color{magenta}\aap},
  \href{https://ui.adsabs.harvard.edu/abs/2019A&A...627A..24G}{627, A24}

\bibitem[{{Guer{\c{c}}o} {et~al.}(2019){Guer{\c{c}}o}, {Cunha}, {Smith},
  {Hayes}, {Abia}, {Lambert}, {J{\"o}nsson}, \& {Ryde}}]{Guer2019}
{Guer{\c{c}}o}, R., {Cunha}, K., {Smith}, V.~V., {et~al.} 2019,
  \href{http://dx.doi.org/10.3847/1538-4357/ab45f1}{\color{magenta}\apj},
  \href{https://ui.adsabs.harvard.edu/abs/2019ApJ...885..139G}{885, 139}

\bibitem[{{Guer{\c{c}}o} {et~al.}(2022{\natexlab{a}}){Guer{\c{c}}o},
  {Ram{\'\i}rez}, {Cunha}, {Smith}, {Prantzos}, {Sellgren}, \&
  {Daflon}}]{Guer2022}
{Guer{\c{c}}o}, R., {Ram{\'\i}rez}, S., {Cunha}, K., {et~al.}
  2022{\natexlab{a}},
  \href{http://dx.doi.org/10.3847/1538-4357/ac5c55}{\color{magenta}\apj},
  \href{https://ui.adsabs.harvard.edu/abs/2022ApJ...929...24G}{929, 24}

\bibitem[{{Guer{\c{c}}o} {et~al.}(2022{\natexlab{b}}){Guer{\c{c}}o}, {Smith},
  {Cunha}, {Ekstr{\"o}m}, {Abia}, {Plez}, {Meynet}, {Ramirez}, {Prantzos},
  {Sellgren}, {Hayes}, \& {Majewski}}]{RSGF2022}
{Guer{\c{c}}o}, R., {Smith}, V.~V., {Cunha}, K., {et~al.} 2022{\natexlab{b}},
  \href{http://dx.doi.org/10.1093/mnras/stac2393}{\color{magenta}\mnras},
  \href{https://ui.adsabs.harvard.edu/abs/2022MNRAS.516.2801G}{516, 2801}

\bibitem[{{He} {et~al.}(2018){He}, {Lombardo}, {Dell'Aquila}, {Xu}, {Zhang}, \&
  {Liu}}]{F19palphaO162018}
{He}, J.-J., {Lombardo}, I., {Dell'Aquila}, D., {et~al.} 2018,
  \href{http://dx.doi.org/10.1088/1674-1137/42/1/015001}{\color{magenta}Chinese
  Physics C}, \href{https://ui.adsabs.harvard.edu/abs/2018ChPhC..42a5001H}{42,
  015001}

\bibitem[{{Heger} {et~al.}(2005){Heger}, {Kolbe}, {Haxton}, {Langanke},
  {Mart{\'\i}nez-Pinedo}, \& {Woosley}}]{Heger2005}
{Heger}, A., {Kolbe}, E., {Haxton}, W.~C., {et~al.} 2005,
  \href{http://dx.doi.org/10.1016/j.physletb.2004.12.017}{\color{magenta}Physics
  Letters B}, \href{https://ui.adsabs.harvard.edu/abs/2005PhLB..606..258H}{606,
  258}

\bibitem[{{Hirai} {et~al.}(2021){Hirai}, {Fujii}, \& {Saitoh}}]{Hirai2021}
{Hirai}, Y., {Fujii}, M.~S., \& {Saitoh}, T.~R. 2021,
  \href{http://dx.doi.org/10.1093/pasj/psab038}{\color{magenta}\pasj},
  \href{https://ui.adsabs.harvard.edu/abs/2021PASJ...73.1036H}{73, 1036}

\bibitem[{{Hirai} {et~al.}(2019){Hirai}, {Wanajo}, \& {Saitoh}}]{Hirai2019}
{Hirai}, Y., {Wanajo}, S., \& {Saitoh}, T.~R. 2019,
  \href{http://dx.doi.org/10.3847/1538-4357/ab4654}{\color{magenta}\apj},
  \href{https://ui.adsabs.harvard.edu/abs/2019ApJ...885...33H}{885, 33}

\bibitem[{{Iliadis} {et~al.}(2010){Iliadis}, {Longland}, {Champagne}, \&
  {Coc}}]{Iliadis2010}
{Iliadis}, C., {Longland}, R., {Champagne}, A.~E., \& {Coc}, A. 2010,
  \href{http://dx.doi.org/10.1016/j.nuclphysa.2010.04.012}{\color{magenta}\nphysa},
  \href{https://ui.adsabs.harvard.edu/abs/2010NuPhA.841..323I}{841, 323}

\bibitem[{{Indelicato} {et~al.}(2017){Indelicato}, {La Cognata}, {Spitaleri},
  {Burjan}, {Cherubini}, {Gulino}, {Hayakawa}, {Hons}, {Kroha}, {Lamia},
  {Mazzocco}, {Mrazek}, {Pizzone}, {Romano}, {Strano}, {Torresi}, \&
  {Tumino}}]{F19palphaO162017}
{Indelicato}, I., {La Cognata}, M., {Spitaleri}, C., {et~al.} 2017,
  \href{http://dx.doi.org/10.3847/1538-4357/aa7de7}{\color{magenta}\apj},
  \href{https://ui.adsabs.harvard.edu/abs/2017ApJ...845...19I}{845, 19}

\bibitem[{{J{\"o}nsson} {et~al.}(2017){J{\"o}nsson}, {Ryde}, {Spitoni},
  {Matteucci}, {Cunha}, {Smith}, {Hinkle}, \& {Schultheis}}]{Jonsson2017}
{J{\"o}nsson}, H., {Ryde}, N., {Spitoni}, E., {et~al.} 2017,
  \href{http://dx.doi.org/10.3847/1538-4357/835/1/50}{\color{magenta}\apj},
  \href{https://ui.adsabs.harvard.edu/abs/2017ApJ...835...50J}{835, 50}

\bibitem[{{Jorissen} {et~al.}(1992){Jorissen}, {Smith}, \&
  {Lambert}}]{Jorissen1992}
{Jorissen}, A., {Smith}, V.~V., \& {Lambert}, D.~L. 1992, \aap,
  \href{https://ui.adsabs.harvard.edu/abs/1992A&A...261..164J}{261, 164}

\bibitem[{{Jos{\'e}}(2012)}]{Jose2012}
{Jos{\'e}}, J. 2012, Bulletin of the Astronomical Society of India,
  \href{https://ui.adsabs.harvard.edu/abs/2012BASI...40..443J}{40, 443}

\bibitem[{{Karakas}(2010)}]{Kara2010}
{Karakas}, A.~I. 2010,
  \href{http://dx.doi.org/10.1111/j.1365-2966.2009.16198.x}{\color{magenta}\mnras},
  \href{https://ui.adsabs.harvard.edu/abs/2010MNRAS.403.1413K}{403, 1413}

\bibitem[{{Karakas} \& {Lattanzio}(2014)}]{Karakas2014}
{Karakas}, A.~I. \& {Lattanzio}, J.~C. 2014,
  \href{http://dx.doi.org/10.1017/pasa.2014.21}{\color{magenta}\pasa},
  \href{https://ui.adsabs.harvard.edu/abs/2014PASA...31...30K}{31, e030}

\bibitem[{{Karakas} {et~al.}(2018){Karakas}, {Lugaro}, {Carlos}, {Cseh},
  {Kamath}, \& {Garc{\'\i}a-Hern{\'a}ndez}}]{Karakas2018}
{Karakas}, A.~I., {Lugaro}, M., {Carlos}, M., {et~al.} 2018,
  \href{http://dx.doi.org/10.1093/mnras/sty625}{\color{magenta}\mnras},
  \href{https://ui.adsabs.harvard.edu/abs/2018MNRAS.477..421K}{477, 421}

\bibitem[{{Kobayashi} {et~al.}(2023){Kobayashi}, {Mandel}, {Belczynski},
  {Goriely}, {Janka}, {Just}, {Ruiter}, {Vanbeveren}, {Kruckow}, {Briel},
  {Eldridge}, \& {Stanway}}]{Koba2023}
{Kobayashi}, C., {Mandel}, I., {Belczynski}, K., {et~al.} 2023,
  \href{http://dx.doi.org/10.3847/2041-8213/acad82}{\color{magenta}\apjl},
  \href{https://ui.adsabs.harvard.edu/abs/2023ApJ...943L..12K}{943, L12}

\bibitem[{{Kroupa}(2001)}]{Kroupa2001}
{Kroupa}, P. 2001,
  \href{http://dx.doi.org/10.1046/j.1365-8711.2001.04022.x}{\color{magenta}\mnras},
  \href{https://ui.adsabs.harvard.edu/abs/2001MNRAS.322..231K}{322, 231}

\bibitem[{{Li} {et~al.}(2013){Li}, {Ludwig}, {Caffau}, {Christlieb}, \&
  {Zhao}}]{Li2013}
{Li}, H.~N., {Ludwig}, H.~G., {Caffau}, E., {Christlieb}, N., \& {Zhao}, G.
  2013,
  \href{http://dx.doi.org/10.1088/0004-637X/765/1/51}{\color{magenta}\apj},
  \href{https://ui.adsabs.harvard.edu/abs/2013ApJ...765...51L}{765, 51}

\bibitem[{{Limongi} \& {Chieffi}(2012)}]{Limongi2012}
{Limongi}, M. \& {Chieffi}, A. 2012,
  \href{http://dx.doi.org/10.1088/0067-0049/199/2/38}{\color{magenta}\apjs},
  \href{https://ui.adsabs.harvard.edu/abs/2012ApJS..199...38L}{199, 38}

\bibitem[{{Limongi} \& {Chieffi}(2018)}]{Limongi2018}
{Limongi}, M. \& {Chieffi}, A. 2018,
  \href{http://dx.doi.org/10.3847/1538-4365/aacb24}{\color{magenta}\apjs},
  \href{https://ui.adsabs.harvard.edu/abs/2018ApJS..237...13L}{237, 13}

\bibitem[{{Longland} {et~al.}(2011){Longland}, {Lor{\'e}n-Aguilar}, {Jos{\'e}},
  {Garc{\'\i}a-Berro}, {Althaus}, \& {Isern}}]{Longland2011}
{Longland}, R., {Lor{\'e}n-Aguilar}, P., {Jos{\'e}}, J., {et~al.} 2011,
  \href{http://dx.doi.org/10.1088/2041-8205/737/2/L34}{\color{magenta}\apjl},
  \href{https://ui.adsabs.harvard.edu/abs/2011ApJ...737L..34L}{737, L34}

\bibitem[{{Lucatello} {et~al.}(2011){Lucatello}, {Masseron}, {Johnson},
  {Pignatari}, \& {Herwig}}]{Lucatello2011}
{Lucatello}, S., {Masseron}, T., {Johnson}, J.~A., {Pignatari}, M., \&
  {Herwig}, F. 2011,
  \href{http://dx.doi.org/10.1088/0004-637X/729/1/40}{\color{magenta}\apj},
  \href{https://ui.adsabs.harvard.edu/abs/2011ApJ...729...40L}{729, 40}

\bibitem[{{Lugaro} {et~al.}(2008){Lugaro}, {de Mink}, {Izzard}, {Campbell},
  {Karakas}, {Cristallo}, {Pols}, {Lattanzio}, {Straniero}, {Gallino}, \&
  {Beers}}]{Lugaro2008}
{Lugaro}, M., {de Mink}, S.~E., {Izzard}, R.~G., {et~al.} 2008,
  \href{http://dx.doi.org/10.1051/0004-6361:20079169}{\color{magenta}\aap},
  \href{https://ui.adsabs.harvard.edu/abs/2008A&A...484L..27L}{484, L27}

\bibitem[{{Maeder}(1992)}]{Maeder1992_baryonic}
{Maeder}, A. 1992, \aap,
  \href{https://ui.adsabs.harvard.edu/abs/1992A&A...264..105M}{264, 105}

\bibitem[{{Maeder}(1997)}]{Maeder1997}
{Maeder}, A. 1997, \aap,
  \href{https://ui.adsabs.harvard.edu/abs/1997A&A...321..134M}{321, 134}

\bibitem[{{Maeder} \& {Meynet}(2000)}]{OG2000}
{Maeder}, A. \& {Meynet}, G. 2000,
  \href{http://dx.doi.org/10.48550/arXiv.astro-ph/0006405}{\color{magenta}\aap},
  \href{https://ui.adsabs.harvard.edu/abs/2000A&A...361..159M}{361, 159}

\bibitem[{{Martinet} {et~al.}(2023){Martinet}, {Meynet}, {Ekstr{\"o}m},
  {Georgy}, \& {Hirschi}}]{Martinet2023}
{Martinet}, S., {Meynet}, G., {Ekstr{\"o}m}, S., {Georgy}, C., \& {Hirschi}, R.
  2023,
  \href{http://dx.doi.org/10.1051/0004-6361/202347514}{\color{magenta}\aap},
  \href{https://ui.adsabs.harvard.edu/abs/2023A&A...679A.137M}{679, A137}

\bibitem[{{Martinet} {et~al.}(2021){Martinet}, {Meynet}, {Ekstr{\"o}m},
  {Sim{\'o}n-D{\'\i}az}, {Holgado}, {Castro}, {Georgy}, {Eggenberger},
  {Buldgen}, {Salmon}, {Hirschi}, {Groh}, {Farrell}, \&
  {Murphy}}]{Martinet2021}
{Martinet}, S., {Meynet}, G., {Ekstr{\"o}m}, S., {et~al.} 2021,
  \href{http://dx.doi.org/10.1051/0004-6361/202039426}{\color{magenta}\aap},
  \href{https://ui.adsabs.harvard.edu/abs/2021A&A...648A.126M}{648, A126}

\bibitem[{{Meynet} \& {Arnould}(2000)}]{GMMA2000}
{Meynet}, G. \& {Arnould}, M. 2000,
  \href{http://dx.doi.org/10.48550/arXiv.astro-ph/0001170}{\color{magenta}\aap},
  \href{https://ui.adsabs.harvard.edu/abs/2000A&A...355..176M}{355, 176}

\bibitem[{{Meynet} \& {Maeder}(2003)}]{MM2003}
{Meynet}, G. \& {Maeder}, A. 2003,
  \href{http://dx.doi.org/10.1051/0004-6361:20030512}{\color{magenta}\aap},
  \href{https://ui.adsabs.harvard.edu/abs/2003A&A...404..975M}{404, 975}

\bibitem[{{Mowlavi} {et~al.}(1996){Mowlavi}, {Jorissen}, \&
  {Arnould}}]{Mowlavi1996}
{Mowlavi}, N., {Jorissen}, A., \& {Arnould}, M. 1996,
  \href{http://dx.doi.org/10.48550/arXiv.astro-ph/9602138}{\color{magenta}\aap},
  \href{https://ui.adsabs.harvard.edu/abs/1996A&A...311..803M}{311, 803}

\bibitem[{{Mura-Guzm{\'a}n} {et~al.}(2020){Mura-Guzm{\'a}n}, {Yong}, {Abate},
  {Karakas}, {Kobayashi}, {Oh}, {Chun}, \& {Mace}}]{Mura2020}
{Mura-Guzm{\'a}n}, A., {Yong}, D., {Abate}, C., {et~al.} 2020,
  \href{http://dx.doi.org/10.1093/mnras/staa2610}{\color{magenta}\mnras},
  \href{https://ui.adsabs.harvard.edu/abs/2020MNRAS.498.3549M}{498, 3549}

\bibitem[{{Murphy} {et~al.}(2020){Murphy}, {Groh}, {Ekstr{\"o}m}, {Meynet},
  {Pezzotti}, {Georgy}, {Choplin}, {Eggenberger}, {Farrell}, {Haemmerl{\'e}},
  {Hirschi}, {Maeder}, \& {Martinet}}]{Murphy2021}
{Murphy}, L.~J., {Groh}, J.~H., {Ekstr{\"o}m}, S., {et~al.} 2020,
  \href{https://ui.adsabs.harvard.edu/abs/2020MNRAS.tmp.3581M}{\href{http://dx.doi.org/10.1093/mnras/staa3803}{\color{magenta}\mnras}}

\bibitem[{{Nandal} {et~al.}(2024){Nandal}, {Sibony}, \& {Tsiatsiou}}]{DYS2024}
{Nandal}, D., {Sibony}, Y., \& {Tsiatsiou}, S. 2024,
  \href{http://dx.doi.org/10.1051/0004-6361/202348866}{\color{magenta}\aap},
  \href{https://ui.adsabs.harvard.edu/abs/2024A&A...688A.142N}{688, A142}

\bibitem[{{Nugis} \& {Lamers}(2000)}]{Nugis2000}
{Nugis}, T. \& {Lamers}, H.~J.~G.~L.~M. 2000, \aap,
  \href{https://ui.adsabs.harvard.edu/abs/2000A&A...360..227N}{360, 227}

\bibitem[{{Obergaulinger} \& {Aloy}(2020)}]{Ober2020}
{Obergaulinger}, M. \& {Aloy}, M.~{\'A}. 2020,
  \href{http://dx.doi.org/10.1093/mnras/staa096}{\color{magenta}\mnras},
  \href{https://ui.adsabs.harvard.edu/abs/2020MNRAS.492.4613O}{492, 4613}

\bibitem[{{Obergaulinger} {et~al.}(2014){Obergaulinger}, {Janka}, \&
  {Aloy}}]{Ober2014}
{Obergaulinger}, M., {Janka}, H.~T., \& {Aloy}, M.~A. 2014,
  \href{http://dx.doi.org/10.1093/mnras/stu1969}{\color{magenta}\mnras},
  \href{https://ui.adsabs.harvard.edu/abs/2014MNRAS.445.3169O}{445, 3169}

\bibitem[{{Olive} \& {Vangioni}(2019)}]{Olive2019}
{Olive}, K.~A. \& {Vangioni}, E. 2019,
  \href{http://dx.doi.org/10.1093/mnras/stz2893}{\color{magenta}\mnras},
  \href{https://ui.adsabs.harvard.edu/abs/2019MNRAS.490.4307O}{490, 4307}

\bibitem[{{Otsuka} \& {Hyung}(2020)}]{Otsuka2020}
{Otsuka}, M. \& {Hyung}, S. 2020,
  \href{http://dx.doi.org/10.1093/mnras/stz3147}{\color{magenta}\mnras},
  \href{https://ui.adsabs.harvard.edu/abs/2020MNRAS.491.2959O}{491, 2959}

\bibitem[{{Otsuka} {et~al.}(2015){Otsuka}, {Hyung}, \& {Tajitsu}}]{Otsuka2015}
{Otsuka}, M., {Hyung}, S., \& {Tajitsu}, A. 2015,
  \href{http://dx.doi.org/10.1088/0067-0049/217/2/22}{\color{magenta}\apjs},
  \href{https://ui.adsabs.harvard.edu/abs/2015ApJS..217...22O}{217, 22}

\bibitem[{{Otsuka} {et~al.}(2008){Otsuka}, {Izumiura}, {Tajitsu}, \&
  {Hyung}}]{Otsuka2008}
{Otsuka}, M., {Izumiura}, H., {Tajitsu}, A., \& {Hyung}, S. 2008,
  \href{http://dx.doi.org/10.1086/591147}{\color{magenta}\apjl},
  \href{https://ui.adsabs.harvard.edu/abs/2008ApJ...682L.105O}{682, L105}

\bibitem[{{Palacios} {et~al.}(2005{\natexlab{a}}){Palacios}, {Arnould}, \&
  {Meynet}}]{Pala2005}
{Palacios}, A., {Arnould}, M., \& {Meynet}, G. 2005{\natexlab{a}},
  \href{http://dx.doi.org/10.1051/0004-6361:20053323}{\color{magenta}\aap},
  \href{https://ui.adsabs.harvard.edu/abs/2005A&A...443..243P}{443, 243}

\bibitem[{{Palacios} {et~al.}(2005{\natexlab{b}}){Palacios}, {Meynet},
  {Vuissoz}, {Kn{\"o}dlseder}, {Schaerer}, {Cervi{\~n}o}, \&
  {Mowlavi}}]{Palacios2005}
{Palacios}, A., {Meynet}, G., {Vuissoz}, C., {et~al.} 2005{\natexlab{b}},
  \href{http://dx.doi.org/10.1051/0004-6361:20041757}{\color{magenta}\aap},
  \href{https://ui.adsabs.harvard.edu/abs/2005A&A...429..613P}{429, 613}

\bibitem[{{Patton} \& {Sukhbold}(2020)}]{Patton2020}
{Patton}, R.~A. \& {Sukhbold}, T. 2020,
  \href{http://dx.doi.org/10.1093/mnras/staa3029}{\color{magenta}\mnras},
  \href{https://ui.adsabs.harvard.edu/abs/2020MNRAS.499.2803P}{499, 2803}

\bibitem[{{Pignatari} {et~al.}(2016){Pignatari}, {Herwig}, {Hirschi},
  {Bennett}, {Rockefeller}, {Fryer}, {Timmes}, {Ritter}, {Heger}, {Jones},
  {Battino}, {Dotter}, {Trappitsch}, {Diehl}, {Frischknecht}, {Hungerford},
  {Magkotsios}, {Travaglio}, \& {Young}}]{Pignatari2016}
{Pignatari}, M., {Herwig}, F., {Hirschi}, R., {et~al.} 2016,
  \href{http://dx.doi.org/10.3847/0067-0049/225/2/24}{\color{magenta}\apjs},
  \href{https://ui.adsabs.harvard.edu/abs/2016ApJS..225...24P}{225, 24}

\bibitem[{{Prantzos} {et~al.}(2023){Prantzos}, {Abia}, {Chen}, {de Laverny},
  {Recio-Blanco}, {Athanassoula}, {Roberti}, {Vescovi}, {Limongi}, {Chieffi},
  \& {Cristallo}}]{Prantzos2023}
{Prantzos}, N., {Abia}, C., {Chen}, T., {et~al.} 2023,
  \href{http://dx.doi.org/10.1093/mnras/stad1551}{\color{magenta}\mnras},
  \href{https://ui.adsabs.harvard.edu/abs/2023MNRAS.523.2126P}{523, 2126}

\bibitem[{{Prantzos} {et~al.}(2018){Prantzos}, {Abia}, {Limongi}, {Chieffi}, \&
  {Cristallo}}]{Prantzos2018}
{Prantzos}, N., {Abia}, C., {Limongi}, M., {Chieffi}, A., \& {Cristallo}, S.
  2018, \href{http://dx.doi.org/10.1093/mnras/sty316}{\color{magenta}\mnras},
  \href{https://ui.adsabs.harvard.edu/abs/2018MNRAS.476.3432P}{476, 3432}

\bibitem[{{Prantzos} \& {Aubert}(1995)}]{Prantzos1995}
{Prantzos}, N. \& {Aubert}, O. 1995, \aap,
  \href{https://ui.adsabs.harvard.edu/abs/1995A&A...302...69P}{302, 69}

\bibitem[{{Recio-Blanco} {et~al.}(2012){Recio-Blanco}, {de Laverny}, {Worley},
  {Santos}, {Melo}, \& {Israelian}}]{Recio2012}
{Recio-Blanco}, A., {de Laverny}, P., {Worley}, C., {et~al.} 2012,
  \href{http://dx.doi.org/10.1051/0004-6361/201118261}{\color{magenta}\aap},
  \href{https://ui.adsabs.harvard.edu/abs/2012A&A...538A.117R}{538, A117}

\bibitem[{{Renda} {et~al.}(2005){Renda}, {Fenner}, {Gibson}, {Karakas},
  {Lattanzio}, {Campbell}, {Chieffi}, {Cunha}, \& {Smith}}]{Renda2005}
{Renda}, A., {Fenner}, Y., {Gibson}, B.~K., {et~al.} 2005,
  \href{http://dx.doi.org/10.1016/j.nuclphysa.2005.05.058}{\color{magenta}\nphysa},
  \href{https://ui.adsabs.harvard.edu/abs/2005NuPhA.758..324R}{758, 324}

\bibitem[{{Ritter} {et~al.}(2018){Ritter}, {Herwig}, {Jones}, {Pignatari},
  {Fryer}, \& {Hirschi}}]{Ritter2018}
{Ritter}, C., {Herwig}, F., {Jones}, S., {et~al.} 2018,
  \href{http://dx.doi.org/10.1093/mnras/sty1729}{\color{magenta}\mnras},
  \href{https://ui.adsabs.harvard.edu/abs/2018MNRAS.480..538R}{480, 538}

\bibitem[{{Rizzuti} {et~al.}(2024){Rizzuti}, {Hirschi}, {Varma}, {Arnett},
  {Georgy}, {Meakin}, {Moc{\'a}k}, {Murphy}, \& {Rauscher}}]{Rizzuti2024}
{Rizzuti}, F., {Hirschi}, R., {Varma}, V., {et~al.} 2024,
  \href{http://dx.doi.org/10.3390/galaxies12060087}{\color{magenta}Galaxies},
  \href{https://ui.adsabs.harvard.edu/abs/2024Galax..12...87R}{12, 87}

\bibitem[{{Roberti} {et~al.}(2024{\natexlab{a}}){Roberti}, {Limongi}, \&
  {Chieffi}}]{Roberti2024b}
{Roberti}, L., {Limongi}, M., \& {Chieffi}, A. 2024{\natexlab{a}},
  \href{http://dx.doi.org/10.3847/1538-4365/ad391d}{\color{magenta}\apjs},
  \href{https://ui.adsabs.harvard.edu/abs/2024ApJS..272...15R}{272, 15}

\bibitem[{{Roberti} {et~al.}(2024{\natexlab{b}}){Roberti}, {Limongi}, \&
  {Chieffi}}]{Roberti2024a}
{Roberti}, L., {Limongi}, M., \& {Chieffi}, A. 2024{\natexlab{b}},
  \href{http://dx.doi.org/10.3847/1538-4365/ad1686}{\color{magenta}\apjs},
  \href{https://ui.adsabs.harvard.edu/abs/2024ApJS..270...28R}{270, 28}

\bibitem[{{Ryde}(2020)}]{Ryde2020}
{Ryde}, N. 2020,
  \href{http://dx.doi.org/10.1007/s12036-020-09657-4}{\color{magenta}Journal of
  Astrophysics and Astronomy},
  \href{https://ui.adsabs.harvard.edu/abs/2020JApA...41...34R}{41, 34}

\bibitem[{{Saberi} {et~al.}(2022){Saberi}, {Khouri}, {Velilla-Prieto},
  {Fonfr{\'\i}a}, {Vlemmings}, \& {Wedemeyer}}]{Saberi2022}
{Saberi}, M., {Khouri}, T., {Velilla-Prieto}, L., {et~al.} 2022,
  \href{http://dx.doi.org/10.1051/0004-6361/202141704}{\color{magenta}\aap},
  \href{https://ui.adsabs.harvard.edu/abs/2022A&A...663A..54S}{663, A54}

\bibitem[{{Saitoh}(2017)}]{Saitoh2017}
{Saitoh}, T.~R. 2017,
  \href{http://dx.doi.org/10.3847/1538-3881/153/2/85}{\color{magenta}\aj},
  \href{https://ui.adsabs.harvard.edu/abs/2017AJ....153...85S}{153, 85}

\bibitem[{{Sander} {et~al.}(2023){Sander}, {Lefever}, {Poniatowski},
  {Ramachandran}, {Sabhahit}, \& {Vink}}]{Sander2023}
{Sander}, A.~A.~C., {Lefever}, R.~R., {Poniatowski}, L.~G., {et~al.} 2023,
  \href{http://dx.doi.org/10.1051/0004-6361/202245110}{\color{magenta}\aap},
  \href{https://ui.adsabs.harvard.edu/abs/2023A&A...670A..83S}{670, A83}

\bibitem[{{Sander} \& {Vink}(2020)}]{Sander2020}
{Sander}, A. A.~C. \& {Vink}, J.~S. 2020,
  \href{http://dx.doi.org/10.1093/mnras/staa2712}{\color{magenta}\mnras},
  \href{https://ui.adsabs.harvard.edu/abs/2020MNRAS.499..873S}{499, 873}

\bibitem[{{Schuler} {et~al.}(2007){Schuler}, {Cunha}, {Smith}, {Sivarani},
  {Beers}, \& {Lee}}]{Schuler2007}
{Schuler}, S.~C., {Cunha}, K., {Smith}, V.~V., {et~al.} 2007,
  \href{http://dx.doi.org/10.1086/521951}{\color{magenta}\apjl},
  \href{https://ui.adsabs.harvard.edu/abs/2007ApJ...667L..81S}{667, L81}

\bibitem[{{Sibony} {et~al.}(2023){Sibony}, {Georgy}, {Ekstr{\"o}m}, \&
  {Meynet}}]{Sibony2023}
{Sibony}, Y., {Georgy}, C., {Ekstr{\"o}m}, S., \& {Meynet}, G. 2023,
  \href{http://dx.doi.org/10.1051/0004-6361/202346638}{\color{magenta}\aap},
  \href{https://ui.adsabs.harvard.edu/abs/2023A&A...680A.101S}{680, A101}

\bibitem[{{Sibony} {et~al.}(2024){Sibony}, {Shepherd}, {Yusof}, {Hirschi},
  {Chambers}, {Tsiatsiou}, {Nandal}, {Sciarini}, {Moyano}, {B{\'e}trisey},
  {Buldgen}, {Georgy}, {Ekstr{\"o}m}, {Eggenberger}, \& {Meynet}}]{Sibony2024}
{Sibony}, Y., {Shepherd}, K.~G., {Yusof}, N., {et~al.} 2024,
  \href{http://dx.doi.org/10.1051/0004-6361/202450180}{\color{magenta}\aap},
  \href{https://ui.adsabs.harvard.edu/abs/2024A&A...690A..91S}{690, A91}

\bibitem[{{Snow} {et~al.}(2007){Snow}, {Destree}, \& {Jensen}}]{Snow2007}
{Snow}, T.~P., {Destree}, J.~D., \& {Jensen}, A.~G. 2007,
  \href{http://dx.doi.org/10.1086/510187}{\color{magenta}\apj},
  \href{https://ui.adsabs.harvard.edu/abs/2007ApJ...655..285S}{655, 285}

\bibitem[{{Spitoni} {et~al.}(2018){Spitoni}, {Matteucci}, {J{\"o}nsson},
  {Ryde}, \& {Romano}}]{Spitoni2018}
{Spitoni}, E., {Matteucci}, F., {J{\"o}nsson}, H., {Ryde}, N., \& {Romano}, D.
  2018,
  \href{http://dx.doi.org/10.1051/0004-6361/201732092}{\color{magenta}\aap},
  \href{https://ui.adsabs.harvard.edu/abs/2018A&A...612A..16S}{612, A16}

\bibitem[{{Stacy} {et~al.}(2013){Stacy}, {Greif}, {Klessen}, {Bromm}, \&
  {Loeb}}]{Stacy2013a}
{Stacy}, A., {Greif}, T.~H., {Klessen}, R.~S., {Bromm}, V., \& {Loeb}, A. 2013,
  \href{http://dx.doi.org/10.1093/mnras/stt264}{\color{magenta}\mnras},
  \href{https://ui.adsabs.harvard.edu/abs/2013MNRAS.431.1470S}{431, 1470}

\bibitem[{{Stancliffe} {et~al.}(2005){Stancliffe}, {Lugaro}, {Ugalde}, {Tout},
  {G{\"o}rres}, \& {Wiescher}}]{Stancliffe2005}
{Stancliffe}, R.~J., {Lugaro}, M., {Ugalde}, C., {et~al.} 2005,
  \href{http://dx.doi.org/10.1111/j.1365-2966.2005.09081.x}{\color{magenta}\mnras},
  \href{https://ui.adsabs.harvard.edu/abs/2005MNRAS.360..375S}{360, 375}

\bibitem[{{Talon} \& {Zahn}(1997)}]{Talon1997}
{Talon}, S. \& {Zahn}, J.~P. 1997,
  \href{http://dx.doi.org/10.48550/arXiv.astro-ph/9609010}{\color{magenta}\aap},
  \href{https://ui.adsabs.harvard.edu/abs/1997A&A...317..749T}{317, 749}

\bibitem[{{Tinsley}(1980)}]{Tinsley1980}
{Tinsley}, B.~M. 1980,
  \href{http://dx.doi.org/10.48550/arXiv.2203.02041}{\color{magenta}\fcp},
  \href{https://ui.adsabs.harvard.edu/abs/1980FCPh....5..287T}{5, 287}

\bibitem[{{Tsiatsiou} {et~al.}(2024){Tsiatsiou}, {Sibony}, {Nandal},
  {Sciarini}, {Hirai}, {Ekstr{\"o}m}, {Farrell}, {Murphy}, {Choplin},
  {Hirschi}, {Chiappini}, {Liu}, {Bromm}, {Groh}, \& {Meynet}}]{Tsiatsiou2024}
{Tsiatsiou}, S., {Sibony}, Y., {Nandal}, D., {et~al.} 2024,
  \href{http://dx.doi.org/10.1051/0004-6361/202449156}{\color{magenta}\aap},
  \href{https://ui.adsabs.harvard.edu/abs/2024A&A...687A.307T}{687, A307}

\bibitem[{{Ugalde} {et~al.}(2008){Ugalde}, {Azuma}, {Couture}, {G{\"o}rres},
  {Lee}, {Stech}, {Strandberg}, {Tan}, \& {Wiescher}}]{Ugalde2008}
{Ugalde}, C., {Azuma}, R.~E., {Couture}, A., {et~al.} 2008,
  \href{http://dx.doi.org/10.1103/PhysRevC.77.035801}{\color{magenta}\prc},
  \href{https://ui.adsabs.harvard.edu/abs/2008PhRvC..77c5801U}{77, 035801}

\bibitem[{{Vink} {et~al.}(2001){Vink}, {de Koter}, \& {Lamers}}]{Vink2001}
{Vink}, J.~S., {de Koter}, A., \& {Lamers}, H.~J.~G.~L.~M. 2001,
  \href{http://dx.doi.org/10.1051/0004-6361:20010127}{\color{magenta}\aap},
  \href{http://adsabs.harvard.edu/abs/2001A%26A...369..574V}{369, 574}

\bibitem[{{Wang} {et~al.}(2022){Wang}, {Vartanyan}, {Burrows}, \&
  {Coleman}}]{Wang2022}
{Wang}, T., {Vartanyan}, D., {Burrows}, A., \& {Coleman}, M. S.~B. 2022,
  \href{http://dx.doi.org/10.1093/mnras/stac2691}{\color{magenta}\mnras},
  \href{https://ui.adsabs.harvard.edu/abs/2022MNRAS.517..543W}{517, 543}

\bibitem[{{Werner} {et~al.}(2005){Werner}, {Rauch}, \& {Kruk}}]{Werner2005}
{Werner}, K., {Rauch}, T., \& {Kruk}, J.~W. 2005,
  \href{http://dx.doi.org/10.1051/0004-6361:20042258}{\color{magenta}\aap},
  \href{https://ui.adsabs.harvard.edu/abs/2005A&A...433..641W}{433, 641}

\bibitem[{{Werner} {et~al.}(2015){Werner}, {Rauch}, \& {Kruk}}]{Werner2015}
{Werner}, K., {Rauch}, T., \& {Kruk}, J.~W. 2015,
  \href{http://dx.doi.org/10.1051/0004-6361/201526842}{\color{magenta}\aap},
  \href{https://ui.adsabs.harvard.edu/abs/2015A&A...582A..94W}{582, A94}

\bibitem[{{Werner} {et~al.}(2016){Werner}, {Rauch}, \& {Kruk}}]{Werner2016}
{Werner}, K., {Rauch}, T., \& {Kruk}, J.~W. 2016,
  \href{http://dx.doi.org/10.1051/0004-6361/201628892}{\color{magenta}\aap},
  \href{https://ui.adsabs.harvard.edu/abs/2016A&A...593A.104W}{593, A104}

\bibitem[{Williams {et~al.}(2021)Williams, Adsley, Davids, Greife, Hutcheon,
  Karpesky, Lennarz, Lovely, \& Ruiz}]{Williams2021_rates}
Williams, M., Adsley, P., Davids, B., {et~al.} 2021,
  \href{http://dx.doi.org/10.1103/PhysRevC.103.055805}{\color{magenta}Phys.
  Rev. C}, 103, 055805

\bibitem[{{Womack} {et~al.}(2023){Womack}, {Vincenzo}, {Gibson},
  {C{\^o}t{\'e}}, {Pignatari}, {Brinkman}, {Ventura}, \&
  {Karakas}}]{Womack2023}
{Womack}, K.~A., {Vincenzo}, F., {Gibson}, B.~K., {et~al.} 2023,
  \href{http://dx.doi.org/10.1093/mnras/stac3180}{\color{magenta}\mnras},
  \href{https://ui.adsabs.harvard.edu/abs/2023MNRAS.518.1543W}{518, 1543}

\bibitem[{{Woosley} {et~al.}(2002){Woosley}, {Heger}, \&
  {Weaver}}]{Woosley2002}
{Woosley}, S.~E., {Heger}, A., \& {Weaver}, T.~A. 2002,
  \href{http://dx.doi.org/10.1103/RevModPhys.74.1015}{\color{magenta}Reviews of
  Modern Physics},
  \href{https://ui.adsabs.harvard.edu/abs/2002RvMP...74.1015W}{74, 1015}

\bibitem[{{Xu} {et~al.}(2013){Xu}, {Takahashi}, {Goriely}, {Arnould}, {Ohta},
  \& {Utsunomiya}}]{Xu2013}
{Xu}, Y., {Takahashi}, K., {Goriely}, S., {et~al.} 2013,
  \href{http://dx.doi.org/10.1016/j.nuclphysa.2013.09.007}{\color{magenta}\nphysa},
  \href{https://ui.adsabs.harvard.edu/abs/2013NuPhA.918...61X}{918, 61}

\bibitem[{{Yusof} {et~al.}(2022){Yusof}, {Hirschi}, {Eggenberger},
  {Ekstr{\"o}m}, {Georgy}, {Sibony}, {Crowther}, {Meynet}, {Kassim}, {Harun},
  {Maeder}, {Groh}, {Farrell}, \& {Murphy}}]{Yusof2022}
{Yusof}, N., {Hirschi}, R., {Eggenberger}, P., {et~al.} 2022,
  \href{http://dx.doi.org/10.1093/mnras/stac230}{\color{magenta}\mnras},
  \href{https://ui.adsabs.harvard.edu/abs/2022MNRAS.511.2814Y}{511, 2814}

\bibitem[{{Zahn}(1992)}]{Zahn1992}
{Zahn}, J.-P. 1992, \aap,
  \href{http://adsabs.harvard.edu/abs/1992A%26A...265..115Z}{265, 115}

\bibitem[{{Zhang} {et~al.}(2021){Zhang}, {L{\'o}pez}, {Lugaro}, {He}, \&
  {Karakas}}]{F19palphaO162021}
{Zhang}, L.~Y., {L{\'o}pez}, A.~Y., {Lugaro}, M., {He}, J.~J., \& {Karakas},
  A.~I. 2021,
  \href{http://dx.doi.org/10.3847/1538-4357/abef63}{\color{magenta}\apj},
  \href{https://ui.adsabs.harvard.edu/abs/2021ApJ...913...51Z}{913, 51}

\bibitem[{{Ziurys} {et~al.}(1994){Ziurys}, {Apponi}, \& {Phillips}}]{Ziu1994}
{Ziurys}, L.~M., {Apponi}, A.~J., \& {Phillips}, T.~G. 1994,
  \href{http://dx.doi.org/10.1086/174682}{\color{magenta}\apj},
  \href{https://ui.adsabs.harvard.edu/abs/1994ApJ...433..729Z}{433, 729}

\end{thebibliography}



\appendix

\section{Some properties of the stellar models used in the present paper} \label{app:B}

Table~\ref{table:tabini} lists the initial abundance of fluorine used in our models for each metallicity. Figure~\ref{fig:Remnantmass} compares the mass of the remnants using different methods (as mentioned in Sect.~\ref{sec:5}). The significant differences in remnant masses highlight the importance of obtaining more consistent values in future studies.

    \begin{table}[ht]
    \caption{Initial abundances of ${}^{19}$F.}
    \begin{center}
    \begin{tabular}{c|c}
    \hline			
    \hline \noalign{\smallskip}				
    Metallicity	&	$X_{{}^{19}F}$	\\		
    \hline \noalign{\smallskip}			
    $Z=0$	&	0.00	\\ [0.4ex]
    $Z=10^{-5}$	&	8.3854e-11	\\ [0.4ex]
    $Z=0.0004$	&	1.5448e-08	\\ [0.4ex]
    $Z=0.002$	&	7.7241e-08	\\ [0.4ex]
    $Z=0.006$	&	2.3172e-07	\\ [0.4ex]
    $Z=0.014$	&	5.4069e-07	\\ [0.4ex]
    $Z=0.020$	&	7.7241e-07	\\ [0.4ex]
    \hline					
    \end{tabular}
    \end{center}
    \label{table:tabini}
    \end{table}  

    \begin{figure}[ht]
    \includegraphics[width=0.45\textwidth]{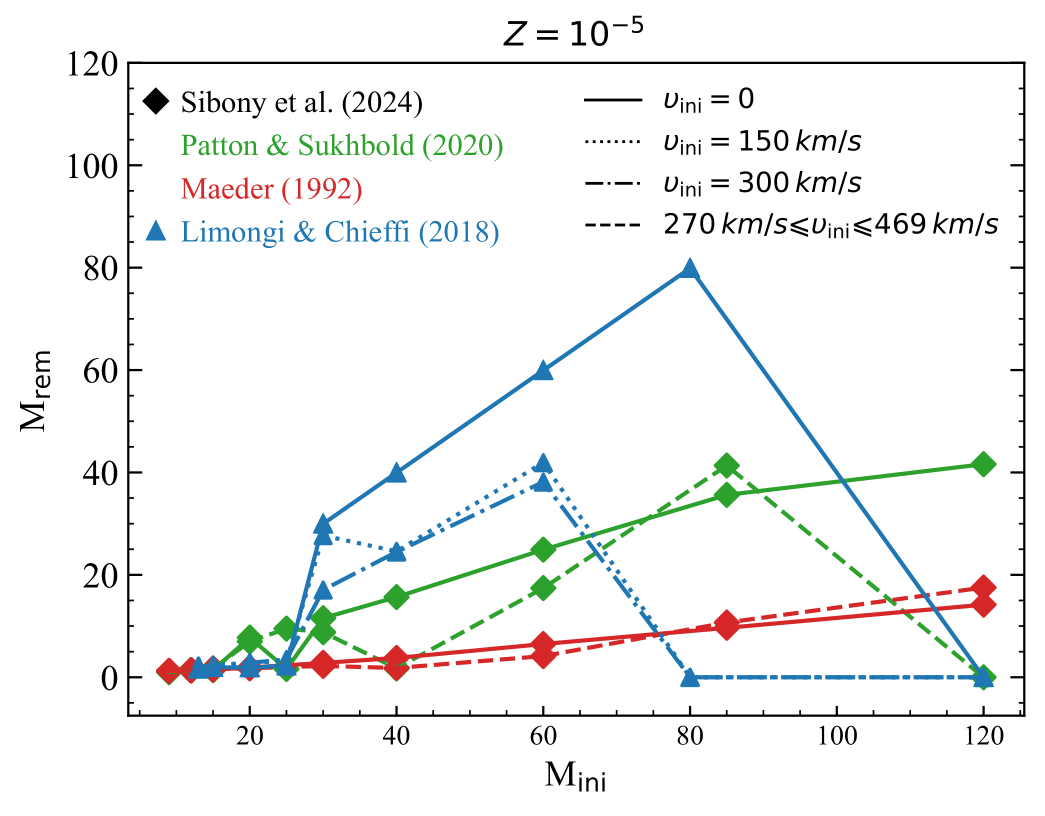}
    \caption{Remnant masses depending on the initial mass for models at EMP metallicity from \citet{Sibony2024} (diamonds) and \citet{Limongi2018} (triangles and blue curves). The remnant masses for \citet{Sibony2024} were computed with two different ways: \citet{Patton2020} (green curves) and \citet{Maeder1992_baryonic} (red curves). The different linestyles present different initial equatorial rotations.}
    \label{fig:Remnantmass}
    \end{figure}

\section{Stellar yields of ${}^{19}$F} \label{app:C}

The wind and SN stellar yields of ${}^{19}$F for both non-rotating and moderately rotating stellar models are provided in units of $10^{-6}$~{\msol}.

    \begin{table}
    \small
    \caption{Values of wind stellar yields of ${}^{19}$F in units of $10^{-6}$~{\msol} for the non-rotating and moderately rotating models.}
    \begin{center}
    \begin{tabular}{c|cccc}
    \hline									
    \hline \noalign{\smallskip}										
    $M_{\rm ini}$ [{\msol}]	&	$Z=0.020$	&	$Z=0.014$	&	$Z=0.006$	&	$Z=0.002$	\\
    \hline \noalign{\smallskip}								
    \multicolumn{5}{c}{$\upsilon_{\rm ini}=0$}									\\
    \hline \noalign{\smallskip}								
    12 & 	$-0.046$ & 	$-0.069$ & 	$-0.003$ & 	$0.0$ \\ [0.4ex]
    15 & 	$-0.037$ & 	$-0.021$ & 	$-0.004$ & 	$-0.001$ \\ [0.4ex]
    20 & 	$-2.249$ & 	$-1.353$ & 	$-0.059$ & 	$-0.006$ \\ [0.4ex]
    25 & 	$-5.466$ & 	$-3.848$ & 	$-0.987$ & 	$0.0$ \\ [0.4ex]
    30 & 	-- & 	-- & 	-- & 	-- \\ [0.4ex]
    32 & 	$-8.067$ & 	$-5.651$ & 	$-2.404$ & 	$-0.058$ \\ [0.4ex]
    40 & 	$-11.16$ & 	$-8.023$ & 	$-3.164$ & 	$-0.186$ \\ [0.4ex]
    60 & 	\textbf{3.072} & 	$-2.936$ & 	$-5.494$ & 	$-0.986$ \\ [0.4ex]
    85 & 	$-1.218$ & 	$-8.821$ & 	$-8.408$ & 	$-2.113$ \\ [0.4ex]
    120 & 	\textbf{11.20} & 	$-12.94$ & 	$-11.36$ & 	$-3.742$ \\ [0.4ex]
    180	& 	--	& 	$-$6.800	& 	$-$29.04	& 	--	\\ [0.4ex]
    250	& 	--	& 	{\bf 68.96}	& 	$-$52.15	& 	--	\\ [0.4ex]
    300	& 	--	& 	$-$0.360	& 	{\bf 742.2}	& 	--	\\ [0.4ex]
	\hline \noalign{\smallskip}							
    \multicolumn{5}{c}{$\upsilon_{\rm ini}=0.4\,\upsilon_{\rm crit}$}		\\
    \hline \noalign{\smallskip}								
    12 & 	$-0.379$ & 	$-0.283$ & 	$-0.051$ & 	$-0.005$ \\ [0.4ex]
    15 & 	$-1.052$ & 	$-0.648$ & 	$-0.060$ & 	$-0.007$ \\ [0.4ex]
    20 & 	$-5.192$ & 	$-3.799$ & 	$-0.597$ & 	$-0.029$ \\ [0.4ex]
    25 & 	$-7.766$ & 	$-5.055$ & 	$-1.919$ & 	$-0.095$ \\ [0.4ex]
    30 & 	-- & 	-- & 	-- & 	-- \\ [0.4ex]
    32 & 	$-7.378$ & 	$-6.870$ & 	$-2.847$ & 	$-0.172$ \\ [0.4ex]
    40 & 	$-15.07$ & 	$-7.436$ & 	$-3.529$ & 	$-0.533$ \\ [0.4ex]
    60 & 	$-15.62$ & 	$-6.690$ & 	$-4.831$ & 	$-1.050$ \\ [0.4ex]
    85 & 	\textbf{5.568} & 	\textbf{0.603} & 	\textbf{18.08} & 	$-2.104$ \\ [0.4ex]
    120 & 	\textbf{0.810} & 	$-17.36$ & 	\textbf{35.90} & 	$-2.115$ \\ [0.4ex]
    180	& 	--	& 	{\bf 45.52}	& 	$-$14.24	& 	--	\\ [0.4ex]
    250	& 	--	& 	--	& 	$-$23.62	& 	--	\\ [0.4ex]
    300	& 	--	& 	$-$104.2	& 	$-$20.66	& 	--	\\ [0.4ex]
    \hline
    \end{tabular}
    \end{center}
    \label{table:yieldswindsF19}
    \end{table}

    \begin{table*}
    \small
    \caption{Values of SN stellar yields of ${}^{19}{\rm F}$ in units of $10^{-6}$~{\msol} for the non-rotating and moderately rotating models. We set values less than $5\times10^{-9}$~{\msol} to zero.} 
    \begin{center}
    \begin{tabular}{c|ccccccc}
    \hline															
    \hline \noalign{\smallskip}							
    $M_{\rm ini}$ [{\msol}]	&	$Z=0.020$	&	$Z=0.014$	&	$Z=0.006$	&	$Z=0.002$	&	$Z=0.0004$	&	$Z=10^{-5}$	&	$Z=0$	\\				
    \hline \noalign{\smallskip}												
    \multicolumn{8}{c}{$\upsilon_{\rm ini}=0$}	\\
    \hline \noalign{\smallskip}														
    9 & 	$-6.055$ & 	$-4.198$ & 	$-1.752$ & 	$-0.592$ & 	$-0.120$ & 	0.0 & 	0.0 \\ [0.4ex]
    12 & 	$-2.458$ & 	$-1.886$ & 	$-0.783$ & 	$-0.278$ & 	$-0.065$ & 	0.0 & 	0.0 \\ [0.4ex]
    15 & 	$-4.080$ & 	$-2.957$ & 	$-1.237$ & 	$-0.429$ & 	$-0.094$ & 	0.0 & 	0.0 \\ [0.4ex]
    20 & 	$-0.518$ & 	$-3.397$ & 	$-2.003$ & 	$-0.311$ & 	$-0.127$ & 	0.0 & 	0.0 \\ [0.4ex]
    25 & 	$-4.374$ & 	$-3.182$ & 	$-1.570$ & 	$-0.958$ & 	$-0.105$ & 	0.0 & 	0.0 \\ [0.4ex]
    30 & 	-- & 	-- & 	-- & 	-- & 	-- & 	0.0 & 	0.0 \\ [0.4ex]
    32 & 	0.0 & 	0.0 & 	0.0 & 	$-0.635$ & 	$-0.159$ & 	-- & 	0.0 \\ [0.4ex]
    40 & 	0.0 & 	0.0 & 	0.0 & 	$-0.953$ & 	$-0.220$ & 	0.0 & 	0.0 \\ [0.4ex]
    60 & 	0.0 & 	0.0 & 	$-0.069$ & 	$-0.611$ & 	\textbf{0.047} & 	\textbf{0.029} & 	0.0 \\ [0.4ex]
    85 & 	0.0 & 	0.0 & 	$-0.181$ & 	$-0.565$ & 	$-0.438$ & 	\textbf{0.069} & 	0.0 \\ [0.4ex]
    120 & 	$-0.132$ & 	$-0.441$ & 	$-2.628$ & 	$-1.114$ & 	-- & 	\textbf{0.005} & 	0.0 \\ [0.4ex]
    \hline \noalign{\smallskip}
    \multicolumn{8}{c}{$\upsilon_{\rm ini}=0.4\,\upsilon_{\rm crit}$} \\
    \hline \noalign{\smallskip}
    9 & 	$-5.931$ & 	$-3.875$ & 	$-1.755$ & 	$-0.587$ & 	$-0.118$ & 	0.0 & 	0.0 \\ [0.4ex]
    12 & 	$-1.864$ & 	$-1.250$ & 	$-0.732$ & 	$-0.286$ & 	\textbf{398.8} & 	\textbf{2283} & 	\textbf{3.622} \\ [0.4ex]
    15 & 	$-1.225$ & 	\textbf{1.440} & 	$-0.953$ & 	$-0.348$ & 	\textbf{7.014} & 	\textbf{60.72} & 	\textbf{0.420} \\ [0.4ex]
    20 & 	\textbf{1.510} & 	0.0 & 	$-0.965$ & 	$-0.498$ & 	$-0.094$ & 	\textbf{10.58} & 	0.0 \\ [0.4ex]
    25 & 	0.0 & 	\textbf{12.58} & 	$-0.273$ & 	$-0.655$ & 	$-0.141$ & 	\textbf{5.982} & 	\textbf{5.297} \\ [0.4ex]
    30 & 	-- & 	-- & 	-- & 	-- & 	-- & 	\textbf{4.379} & 	0.0 \\ [0.4ex]
    32 & 	0.0 & 	0.0 & 	\textbf{108.9} & 	$-0.929$ & 	$-0.215$ & 	-- & 	0.0 \\ [0.4ex]
    40 & 	0.0 & 	0.0 & 	$-0.009$ & 	$-0.517$ & 	\textbf{0.488} & 	\textbf{25.95} & 	0.0 \\ [0.4ex]
    60 & 	$-8.743$ & 	$-0.010$ & 	$-0.122$ & 	$-0.659$ & 	$-0.108$ & 	\textbf{0.791} & 	0.0 \\ [0.4ex]
    85 & 	0.0 & 	$-0.509$ & 	$-0.436$ & 	\textbf{1.373} & 	$-0.214$ & 	\textbf{122.9} & 	0.0 \\ [0.4ex]
    120 & 	$-0.369$ & 	$-0.092$ & 	$-2.247$ & 	$-6.596$ & 	$-1.408$ & 	\textbf{1.094} & 	\textbf{0.021} \\ [0.4ex]
    \hline
    \end{tabular}
    \end{center}
    \label{table:yields}
    \end{table*}

\end{document}